\documentclass{nature}

\usepackage{lineno}

\usepackage{astjnlabbrev-nature}
\usepackage{color}
\usepackage{epsfig}
\usepackage{graphicx,subcaption}
\usepackage{hyperref}

\makeatletter
\let\saved@includegraphics\includegraphics
\AtBeginDocument{\let\includegraphics\saved@includegraphics}
\renewenvironment*{figure}{\@float{figure}}{\end@float}
\makeatother

\usepackage{amsmath,amssymb}
\usepackage{geometry}
\usepackage{pdflscape}
\usepackage{comment}
\usepackage{eqnarray}
\usepackage{helvet}
\usepackage[helvet]{sfmath}
\usepackage{soul}
\usepackage{multibib}

\newcommand{\farcs}{\mbox{\ensuremath{.\!\!^{\prime\prime}}}}

\newcommand{\hbeta}{H{$\beta$}}
\newcommand{\halpha}{H{$\alpha$}}

\newcommand{\CIII}{C{\sevenrm\,III]}}
\newcommand{\MgII}{Mg{\sevenrm\,II}}

\newcommand{\OIII}{[O{\sevenrm\,III}]}

\newcommand{\OIIIb}{[O{\sevenrm\,III}]\,$\lambda$5007}

\newcommand{\NII}{[N{\sevenrm\,II}]}

\newcommand{\obj}{SDSS~J0749$+$2255}
\newcommand{\hst}{{\it HST}}
\newcommand{\chandra}{{\it Chandra}}

 \font\sevenrm=cmr7 scaled 1000

\title{A close quasar pair in a disk-disk galaxy merger at $z=2.17$}

\author{Yu-Ching Chen$^{1,2}$, Xin Liu$^{1,2\thanks{e-mail: xinliuxl@illinois.edu}}$, Adi Foord$^{3}$, Yue Shen$^{1,2}$, Masamune Oguri$^{4,5}$, Nianyi Chen$^{6}$, Tiziana Di Matteo$^{6,7,8}$, Miguel Holgado$^{6}$, Hsiang-Chih Hwang$^{9}$, \& Nadia Zakamska$^{10}$}

\begin{document}

\maketitle

\begin{affiliations}
 \item Department of Astronomy, University of Illinois at Urbana-Champaign, Urbana, IL 61801, USA
 \item National Center for Supercomputing Applications, University of Illinois at Urbana-Champaign, 605 East Springfield Avenue, Champaign, IL 61820, USA
 \item Kavli Institute of Particle Astrophysics and Cosmology, Stanford University, Stanford, CA 94305, USA
 \item Center for Frontier Science, Chiba University, Chiba 263-8522, Japan
 \item Department of Physics, Graduate School of Science, Chiba University, Chiba 263-8522, Japan
 \item McWilliams Center for Cosmology, Department of Physics, Carnegie Mellon University, Pittsburgh, PA 15213, USA
 \item NSF AI Planning Institute for Physics of the Future, Carnegie Mellon University, Pittsburgh, PA 15213, USA
 \item OzGrav-Melbourne, Australian Research Council Centre of Excellence for Gravitational Wave Discovery, Melbourne, VIC 3122, Australia
 \item School of Natural Sciences, Institute for Advanced Study, Princeton, 1 Einstein Drive, NJ 08540, USA
 \item Department of Physics and Astronomy, Johns Hopkins University, Baltimore, MD 21218, USA
\end{affiliations}

\begin{abstract}
Galaxy mergers produce pairs of supermassive black holes (SMBHs), which may be witnessed as dual quasars if both SMBHs are rapidly accreting. The kiloparsec (kpc)-scale separation represents a physical regime sufficiently close for merger-induced effects to be important\cite{Hopkins2008} yet wide enough to be directly resolvable with the facilities currently available. Whereas many kpc-scale dual active galactic nuclei--the low-luminosity counterparts of quasars--have been observed in low-redshift mergers\cite{Bogdanovic2022}, no unambiguous dual quasar is known at cosmic noon ($z\approx2$), the peak of global star formation and quasar activity\cite{Richards2006,Madau2014}. Here we report multiwavelength observations of \obj\ as a kpc-scale, dual-quasar system hosted by a galaxy merger at cosmic noon ($z=2.17$). We discover extended host galaxies associated with the much brighter compact quasar nuclei (separated by 0\farcs46 or 3.8 kpc) and low-surface-brightness tidal features as evidence for galactic interactions. Unlike its low-redshift and low-luminosity counterparts, \obj\ is hosted by massive compact disk-dominated galaxies. The apparent lack of stellar bulges and the fact that \obj\ already follows the local SMBH mass--host stellar mass relation, suggest that at least some SMBHs may have formed before their host stellar bulges. While still at kpc-scale separations where the host-galaxy gravitational potential dominates, the two SMBHs may evolve into a gravitationally bound binary system in around 0.22 Gyr.
\end{abstract}

\obj\ is an optically selected type 1 (that is, broad-line) quasar at $z=2.17$ spectroscopically confirmed by the SDSS legacy survey\cite{Schneider2010}. It was selected as a candidate kpc-scale, dual quasar using Gaia astrometry\cite{HwangShen2020}. Hubble Space Telescope (\hst ) optical dual-band imaging of \obj\ has revealed compact double nuclei with a projected nuclear angular separation of 0\farcs46\ (ref. \cite{Shen2021}). Close compact optical double nuclei could represent a dual quasar, a projected quasar pair in two noninteracting galaxies, a strongly lensed quasar or a star-quasar superposition. Previous very-long-baseline array (VLBA) 15.0 GHz (Ku-band) imaging detected strong radio emissions from both nuclei, ruling out star-quasar superposition\cite{Shen2021}. Whereas morphological differences in the VLBA 15.0 GHz image of the two radio nuclei favor a dual quasar or a premerger projected quasar pair at larger physical separations, strong lensing could not be conclusively ruled out\cite{Shen2021}. 

To unambiguously test the nature of \obj , we have conducted a comprehensive multiwavelength follow-up campaign. The new observations include {\it HST} infrared (IR) F160W ($H$ band) imaging, Keck adaptive optics (AO)-assisted IR ($K_p$ band) imaging, {\it HST}/Space Telescope Imaging Spectrograph (STIS) optical spatially resolved slit spectroscopy, Gemini spatially resolved optical (GMOS) and near infrared (NIR) (GNIRS) slit spectroscopy, {\it Chandra} Advanced CCD Imaging Spectrometer (ACIS)-S X-ray 0.5--8.0 keV imaging spectroscopy, and very-large array ({\it VLA}) A-configuration (A-config) radio dual-band (15 and 6 GHz) imaging. These multiwavelength observations establish \obj\ as a dual quasar hosted by an ongoing galaxy merger at $z=2.17$.

\begin{figure}
\centerline{
\includegraphics[width=1.3\textwidth]{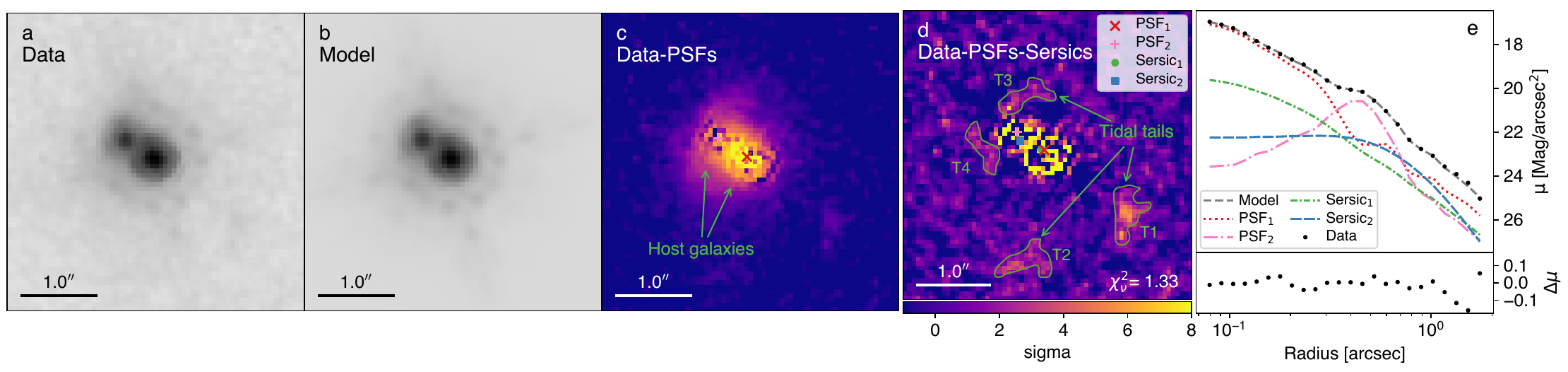}
}
\caption{\textbf{HST imaging. a--e}, HST/WFC3 F160W (H-band) image and two-dimensional structural decomposition results using GALFIT\cite{Peng2010} modeling for \obj. \textbf{a}, Raw HST image. \textbf{b}, Our best-fit model consisting of two PSFs for the central quasars and two S\'{e}rsic models for the two host galaxies. \textbf{c}, Two-PSF-subtracted image highlighting the detection of two host galaxies underlying both quasars in the merger. \textbf{d}, Residual image after subtraction of two PSFs and two S\'{e}rsic models, showing detection of low-surface-brightness tidal tails indicative of ongoing galactic interactions. \textbf{e}, One-dimensional radial profiles of the data and best-fit model and their individual components. Detection of both host galaxies and tidal tails unambiguously confirms the system as a dual quasar rather than a lensed quasar or a premerger projected quasar pair at larger physical separations. In addition to the two S\'{e}rsic models for the two quasar host galaxies, we have also considered the alternative scenario of a single S\'{e}rsic model for a central foreground lens galaxy, which is disfavored by the data based on the $\chi^2_{\nu}$ values.}
\label{fig:hst_galfit}
\end{figure}

First, after we subtract the bright central quasars modeled by two point-spread functions (PSFs), the deep {\it HST} IR image uncovers two extended host galaxies underlying both compact quasar nuclei (Figure \ref{fig:hst_galfit}). In addition, after we further subtract the best axisymmetric models for the extended host galaxies, the final residual image shows low-surface-brightness tidal features indicative of mergers (Figure \ref{fig:hst_galfit}). Detection of these faint tidal features is of great significance (over 10$\sigma$ for each integrated feature; Supplementary Information). Detecting both tidal features and the extended host galaxies unambiguously confirms \obj\ as a dual quasar in a merger rather than a strongly lensed quasar or a premerger projected quasar pair at larger physical separations. Complementary higher-angular-resolution Keck AO-assisted IR imaging independently supports the detection of the extended host galaxies and the nondetection of a central foreground lens galaxy. The alternative scenario of a foreground lens galaxy as a single extended source located between two compact quasar nuclei is quantitatively ruled out by strong lensing mass model tests (Supplementary Information). However, the tidal features seen in the \hst\ image are too faint to detect in the shallower Keck AO image.

\begin{figure}
  \centering
  \includegraphics[width=\textwidth]{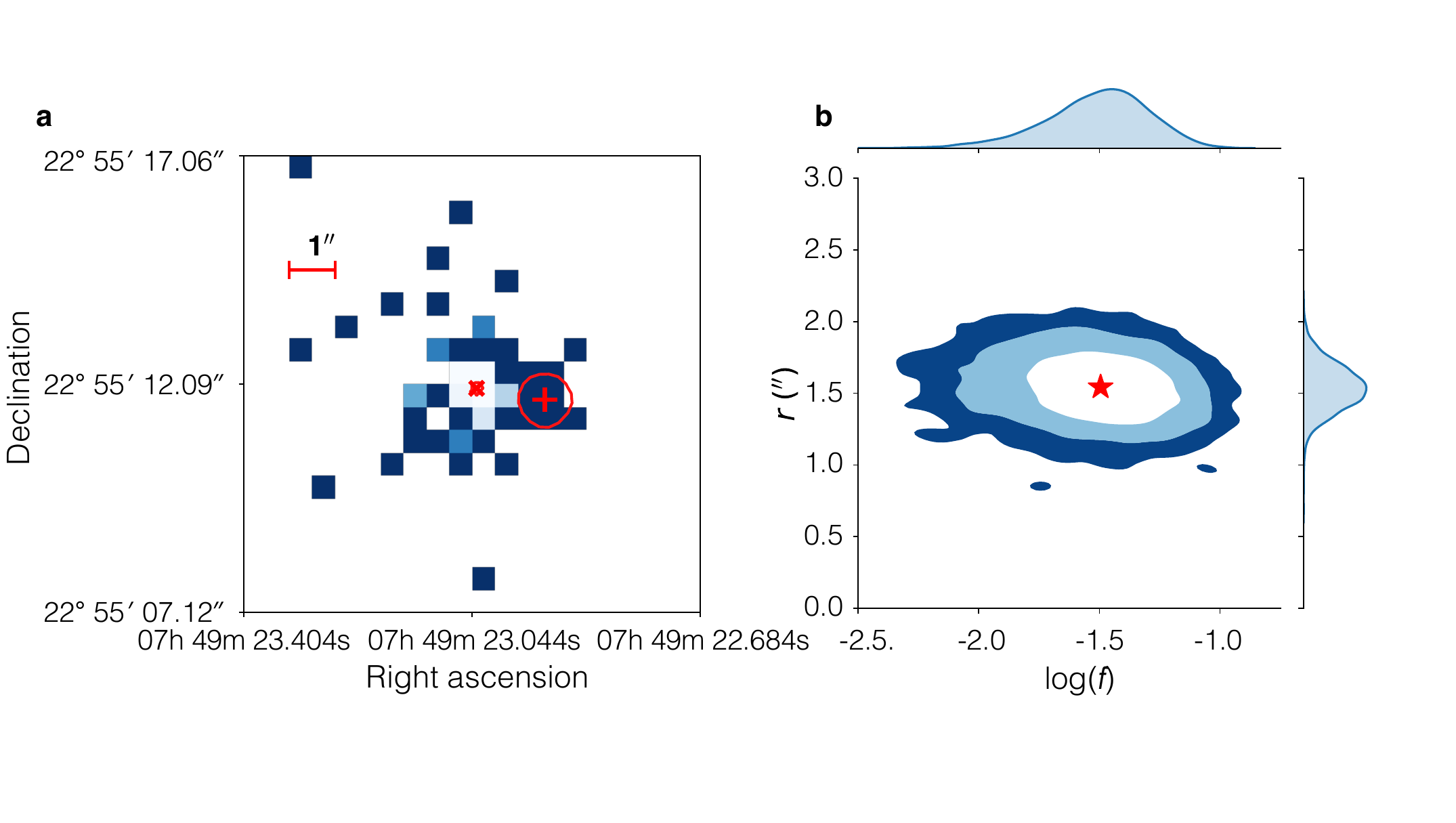}
    \caption{{\bf X-ray imaging. a,b,} \emph{Chandra} ACIS-S X-ray image and best-fit {\tt BAYMAX} results for \obj . \textbf{a}, Binned \chandra\ data (0.5--8.0 keV), centered on the nominal X-ray coordinates of quasar with best-fit location of primary X-ray point source (red ``x") and secondary X-ray point source (red ``+"), with 95\% confidence interval in red contour. \textbf{b}, Joint-posterior distribution of the separation (\emph{r}, in arcseconds) and count ratio (log $f$, ratio between the number of counts associated with secondary versus the primary). Blue contour levels represent 68\% and 95\% confidence interval. We find the best-fit separation $r=1\farcs54_{-0\farcs36}^{+0\farcs39}$ and log$f=-1.43_{-0.30}^{+0.30}$ at the 95\% confidence interval. Although the estimated separation is statistically larger than that estimated from \emph{HST} imaging (0\farcs46), the estimated position angle between the two X-ray point sources ($350_{-24}^{+37}$$^{\circ}$, 95\% confidence interval) is consistent with that of the two resolved cores resolved in the \emph{HST} imaging ($330^{\circ}$). 
    }
    \label{fig:xray}
\end{figure}

Second, \chandra\ ACIS-S detects both quasar nuclei in hard X-rays (2--8 keV) as compact X-ray point sources whose spatial profiles are consistent with the PSF (Figure \ref{fig:xray}). The X-ray positional offsets between the two nuclei are consistent with those of the \hst\ optical nuclei. The optically fainter northeast (NE) nucleus is brighter in the X-rays than the southwest (SW) nucleus. This contradicts naive expectations from a strongly lensed quasar, because gravitational lensing is in general achromatic: the deflection angle of a light ray (and the lensing magnification) does not depend on its wavelength, so consistent flux ratios are expected at different wavelengths from different-lensed images given the same underlying source spectrum. While the wavelength-dependent geometry of the different emission regions may result in chromatic effects\cite{Barnacka2014}, differences in flux ratios between different images at different wavelengths should be minor in a strongly lensed quasar. However, the 2--8 keV X-ray flux ratio between the two nuclei (NE/SW) is 45$^{+41}_{-33}$, which is significantly different from that observed by {\it HST} in the optical (0.229$\pm$0.001 in F475W and 0.284$\pm$0.002 in F814W), disfavoring strong lensing. A caveat is that microlensing can cause such flux ratio anomalies given the different optical and X-ray-emitting region sizes frequently observed\cite{Chang1979,Wambsganss1991,Pooley2006}, although the discrepancy for \obj\ is seen as extreme. The X-ray-to-optical luminosity ratio typically characterized by the spectral slopes $\alpha_{{\rm OX}}$ is 1.46$\pm$0.01 for the NE nucleus and 2.32$\pm$0.13 for the SW nucleus (Supplementary Information). The estimated unabsorbed hard X-ray (2--8 keV) luminosities (assuming a redshift of z = 2.17 for both X-ray point sources) are approximately 10$^{45.29^{+0.05}_{-0.03}}$ erg s$^{-1}$ for the NE nucleus and 10$^{43.62^{+0.32}_{-0.58}}$ erg s$^{-1}$ for the SW nucleus, both greatly exceeding the most X-ray-luminous starburst galaxies\cite{zezas01}. We also find significantly different intrinsic absorbing columns in the two nuclei (4.7$^{+1.3}_{-2.6}\times10^{22}$ cm$^{-2}$ for the NE nucleus and below $10^{20}$ cm$^{-2}$ for the SW nucleus), lending further support to the dual-quasar hypothesis.

\begin{figure}
  \centering
    \includegraphics[width=\textwidth]{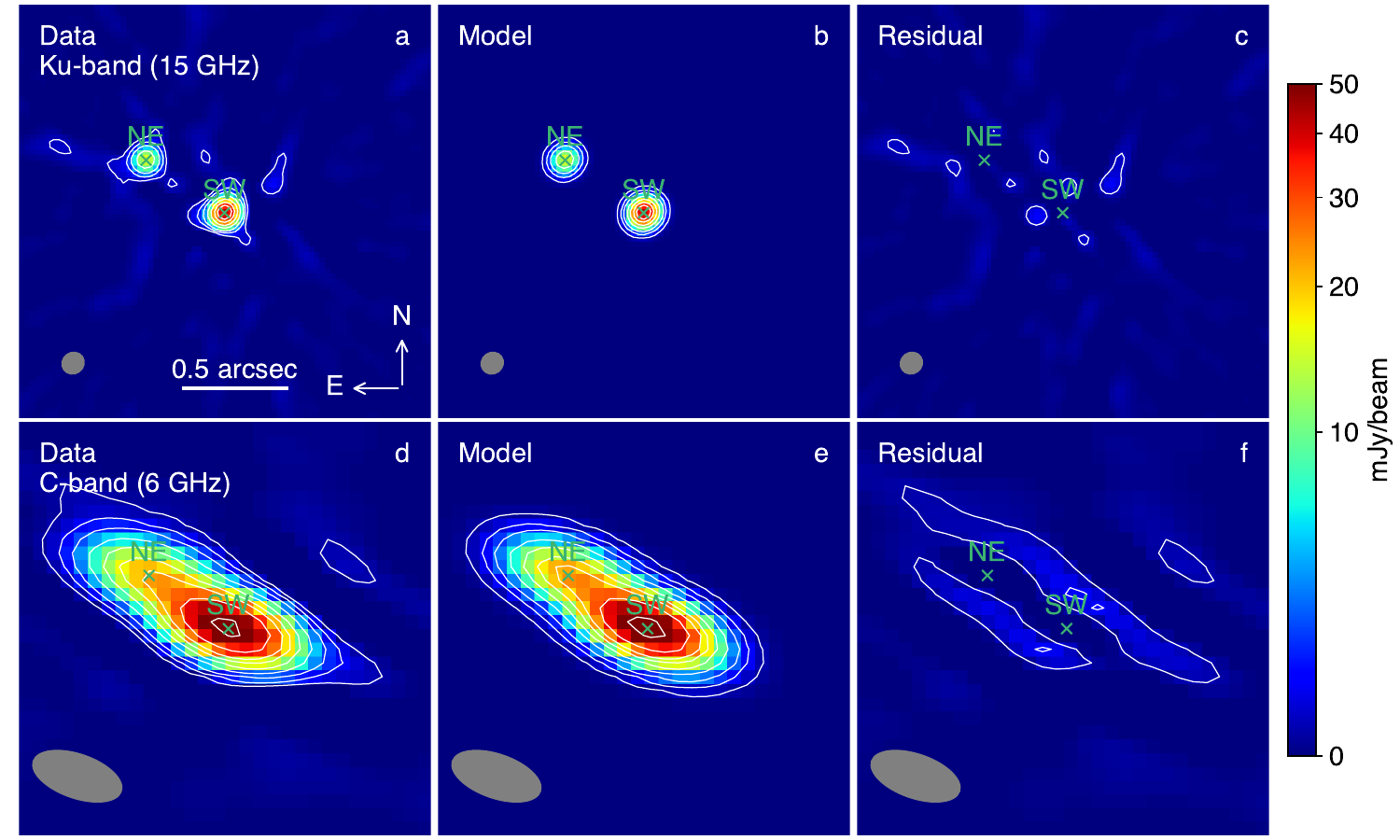}
    \caption{{\bf Radio imaging. a}--\textbf{f},
    VLA A-config Ku-band (\textbf{a}) and C-band (\textbf{d}) continuum images for \obj . \textbf{b},\textbf{c}, The Ku-band model (\textbf{b}) and residual (\textbf{c}) images. \textbf{e},\textbf{f}, The C-band model (\textbf{e}) and residual (\textbf{f}) images. Contours represent 10, 30, 60, 120, 200, 300, 400, 600, and 1,000$\times$1$\sigma$ noise levels. Green crosses denote centroids of the fitted Gaussian components. The synthesized beam sizes are shown in grey in the bottom left-band corner. The restoring beam sizes are 0\farcs11$\times$0\farcs10 (PA=$-54.0^{\circ}$) in the Ku-band and 0\farcs44$\times$0\farcs21 (PA=$70.9^{\circ}$) in the C-band. The rms noise levels are 8.2 $\mu$Jy per beam in the Ku-band and 5.3 $\mu$Jy per beam in the C-band.
    }
    \label{fig:vla}
\end{figure}

Third, both quasar nuclei are detected as compact radio sources by {\it VLA} A-config continuum imaging at 6 and 15 GHz (Figure \ref{fig:vla}). The radio point source centers are consistent with those of the \hst\ optical quasar nuclei. Radio flux ratios between the two nuclei (NE/SW) are 0.379$\pm$0.001 at 6 GHz and 0.351$\pm$0.003 at 15 GHz, which are different (by over $5\sigma$) than those observed by \hst\ in the optical (0.229$\pm$0.001 in F475W and 0.284$\pm$0.002 in F814W), disfavoring strong lensing but with the caveat of flux ratio anomalies caused by microlensing. The radio spectral indices are $-$0.397$\pm$0.004 for the NE nucleus and $-$0.314$\pm$0.002 for the SW nucleus. The radio loudness parameter $R_{6\,cm/2500\,\AA}$, commonly defined as the flux density ratio at the rest-frame 6 cm and that at 2,500 \AA , is 1079$\pm$74 for the NE nucleus and 659$\pm$16 for the SW nucleus, placing both quasars in the radio-loud (that is, $R_{6\,cm/2500\,\AA}$ over 10) population according to the canonical definition\cite{Kellermann1989}.

\begin{figure}
  \centering
 \includegraphics[width=1.0\textwidth]{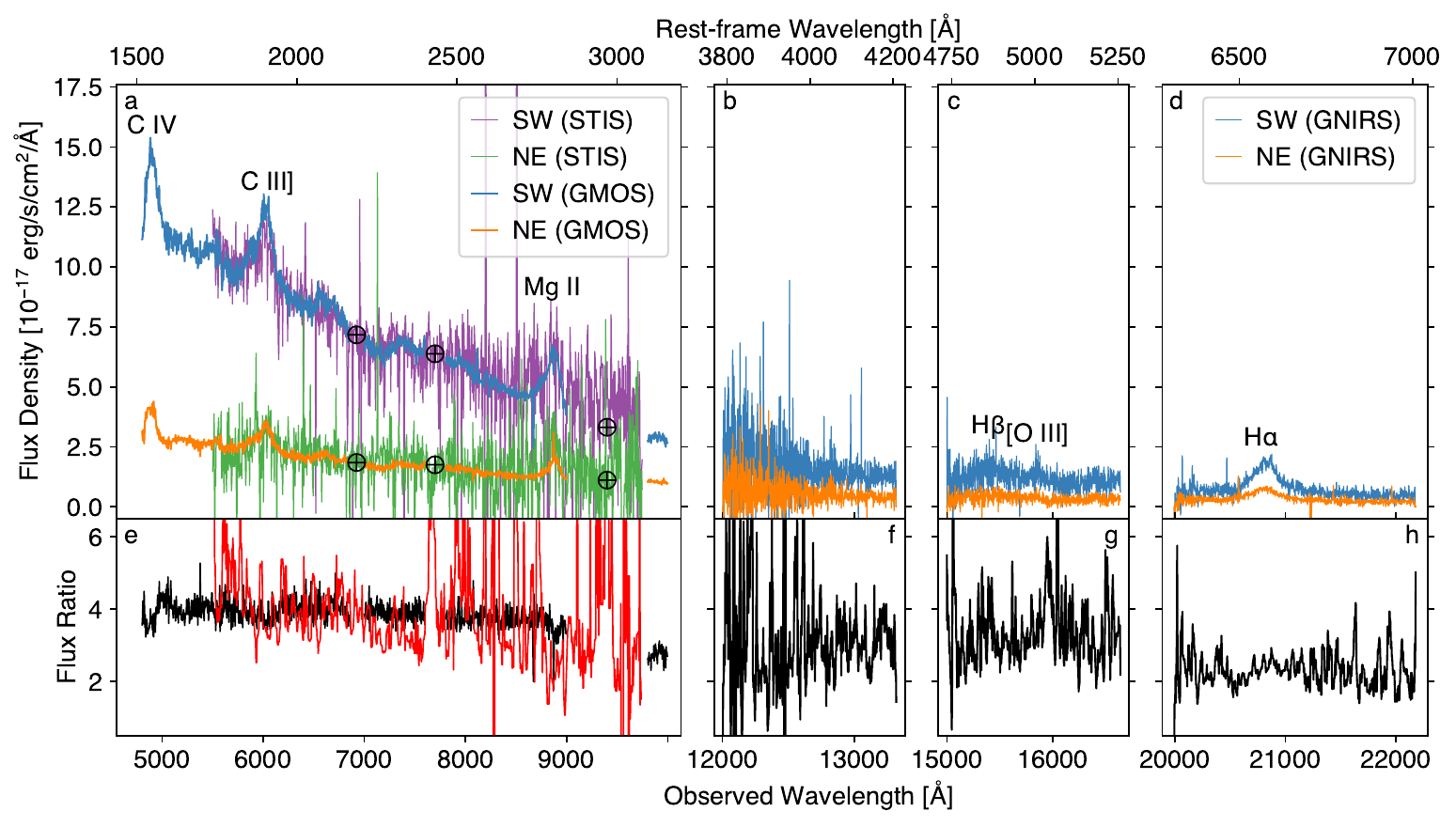}
    \caption{{\bf Optical and NIR spectroscopy. a}--\textbf{d}, Spatially resolved optical and NIR spectroscopy from \hst /STIS, Gemini/GMOS and Gemini/GNIRS for \obj . Orange and cyan curves represent Gemini spectra for the NE and SW nuclei, respectively; green and magenta curves represent \hst\ spectra for the NE and SW nuclei, respectively. Labeled are the major strong emission lines detected. Circled cross symbols denote the telluric absorption region. \textbf{e}--\textbf{h}, Flux ratios between the two nuclei, with the {\it HST}-based flux ratio shown in red and the Gemini-based flux ratio (for both GMOS and GNIRS) in black. The two nuclei have significantly different rest-frame UV continuum spectral slopes and strengths of the broad and narrow emission lines. The discrepancy between STIS- and GMOS-based flux ratios is probably caused by differences in slit coverage and/or observation time. We adopt STIS spectra to estimate the continuum power-law spectral indices, because STIS cleanly separated the two nuclei. 
    }
    \label{fig:OIR_spec}
\end{figure}

Finally, \hst\ spatially resolved optical spectroscopy shows significantly different rest-frame ultraviolet continuum spectral slopes between the two nuclei ($\alpha_{\nu}$ = $-$0.70$\pm$0.24 for the NE nucleus and 0.08$\pm$0.06 for the SW nucleus), independently confirmed by Gemini/GMOS spatially resolved optical slit spectroscopy (Figure \ref{fig:OIR_spec}). However, the GMOS spectra are only marginally resolved for the two nuclei and hence the spectral slope measurements are not as reliable as the HST-based results. The different spectral slopes disfavor a strongly lensed quasar, although strong gravitational lenses may also produce different spectral slopes due to varying reddening along different light paths through the foreground lens galaxy\cite{Sluse2012}. There are noticeable differences in the broad emission lines, but the low spectroscopic data quality cannot conclusively rule out strong lensing based on the line shapes alone. The best-fit systemic redshifts estimated from the optical-IR spectra of the two quasar nuclei suggest a line-of-sight velocity offset of roughly 310$\pm$50 km s$^{-1}$ (Supplementary Table 1), consistent with expectations of typical orbital velocities in galaxy mergers\cite{ContrerasSantos2022} but inconsistent with strong lensing (which would produce identical redshifts of the two quasars).

By modeling the PSF-subtracted, host-galaxy-only, two-band (\hst\ F160W and Keck AO $K_p$) photometry of \obj , we estimate the stellar masses of the two host galaxies at around $10^{11.46^{+0.02}_{-0.26}}M_{\odot}$ and $10^{11.50^{+0.04}_{-0.27}}M_{\odot}$ (with the errors accounting for the 1$\sigma$ statistical and partial systematic uncertainties from surface brightness profile fitting but ignoring additional systematics from population synthesis modeling) for the NE and SW nuclei. The massive hosts appear to be compact and disk dominated (with best-fit S\'{e}rsic indices close to unity; Supplementary Information), typical of galaxies\cite{Shapley2011} and hosts of single quasars\cite{Ding2022a} at $z>2$. The disk dominance of the hosts in \obj\ is dramatically different from the more bulge-dominated characteristics of low-redshift, low-luminosity dual active galactic nucleus hosts\cite{Shangguan2016}. The latter may be more subject to the stabilization effect of stellar bulges preventing more efficient SMBH fueling in the early merging stages, resulting in low-luminosity active galactic nuclei rather than the luminous quasars observed in \obj . 

By jointly modeling the optical-NIR spectrum (Supplementary Information), we estimate the SMBH masses as around $10^{9.1\pm0.4}M_{\odot}$ and $10^{9.2\pm0.4}M_{\odot}$ (with errors representing 1$\sigma$ total uncertainties) for the NE and SW nuclei, respectively. Bolometric luminosities estimated from the optical continua suggest Eddington ratios of about 0.12$\pm$0.03 and 0.25$\pm$0.07 for the NE and SW nuclei, respectively. SMBH masses and host total stellar masses are consistent with the empirical scaling relations observed in both single quasars at $z\approx2$ and local inactive galaxies within uncertainties (Supplementary Information).

\obj\ represents a robust example of a kpc-scale pair of SMBHs in a galaxy merger at cosmic noon. It fills a long-standing gap in the expected population of dual quasars hosted by high-redshift ongoing galaxy mergers. Such systems are expected to be common in hierarchical merger models, which have otherwise been successful in explaining much of the quasar phenomenology\cite{Hopkins2008}. In comparison, the previous redshift record holder of kpc-scale dual quasars hosted by confirmed galaxy mergers was LBQS 0103$-$2753 at $z=0.86$, which was serendipitously discovered\cite{Shields2012}.

A handful of galactic-scale (below 10 kpc) dual quasar candidates were known at $z>1$ (refs.\cite{Inada2008,Schechter2017,Anguita2018,Lemon2018,Lemon2020,Shen2021,Tang2021,Yue2021,Lemon2022}). However, none of these has been confirmed to be hosted by ongoing galaxy mergers--as expected in bona fide dual quasars (Supplementary Information). The establishment of \obj\ as a dual quasar hosted in a galaxy merger conclusively demonstrates that at least some high-redshift, kpc-scale, dual-quasar candidates may indeed be dual quasars rather than gravitational lenses or premerger projected quasar pairs at large separations. On larger (that is, over 10 kpc) scales, it has long been suggested that most wide-separation (3$''$--10$''$) double quasars are quasar pairs rather than gravitational lenses based on statistical arguments\cite{Kochanek1999}. There are approximately tens of known ``binary'' quasars at $z>1$ with projected separations of over 10 kpc\cite{Hennawi2010}, and even a quasar ``quartet''\cite{Hennawi2015}. However, the separations of most of these systems are still too large (approximately over tens of kpc) for the host galaxies of the quasar pair to be in direct galactic interactions. Unlike \obj , none of these has been observed with tidal features as direct evidence for ongoing interactions. 

Despite significant merit and intense theoretical interest, direct evidence for compact binary SMBHs--SMBHs that are gravitationally bound to each other--is still lacking\cite{Bogdanovic2022}. Their separations (typically less than a few parsecs) are too small to resolve beyond the local universe with current facilities. High-redshift, kpc-scale, dual quasars represent promising progenitors of low-redshift compact binary SMBHs. Using a stellar dynamical friction argument, we estimate that the 3.8 kpc SMBH pair in \obj\ may form a compact bound SMBH binary in fewer than approximately 0.22$^{+0.50}_{-0.16}$ Gyr (Supplementary Information). Future infrared integral field unit spectroscopy with the JWST will map the stellar and gas kinematics, enabling better predictions on the subsequent SMBH pair orbital evolution and the impact of dual quasar on their host galaxies. 

The confirmation of \obj\ as a dual quasar provides a proof of concept for combining multiwavelength facilities to discover high-redshift, kpc-scale dual quasars\cite{Shen2021}. The discovery underscores the need for high-resolution deep NIR imaging, which is crucial for detection of merging galaxy hosts and for robustly distinguishing of high-redshift dual quasars from lensed quasars and premerger projected quasar pairs at larger physical separations. Future high-resolution, wide-area deep surveys will uncover a larger sample of high-redshift dual quasars at the peak epoch of galaxy and SMBH assembly.

\bibliographystyle{naturemag}



\begin{addendum}
 \item We thank A. Kemball and A. Gross for helpful discussions on strong lensing. We thank M. Leveille, A. Vick, R. Campbell, R. McGurk, J. Cortes, T. R. Geballe, S. Leggett, A. Nitta, T. Seccull, and H. Medlin for their help with our HST, Keck, Gemini, and VLA observations. Y.-C.C. thanks J. Li and H. Guo for their help with the GALFIT and PyQSOFit codes. 
 M.O. acknowledges support from JSPS KAKENHI grants JP20H00181, JP20H05856, JP22H01260.
 This work is supported by the Heising-Simons Foundation and Research Corporation for Science Advancement and NSF grants AST-2108162 and AST-2206499. 
 This research was supported in part by the National Science Foundation under PHY-1748958. 
 Support for Program number 23700237 was provided by NASA through Chandra Award Number GO2-23099X issued by the Chandra X-ray Observatory Center, which is operated by the Smithsonian Astrophysical Observatory for and on behalf of NASA under contract NAS 8-03060. 
 Support for Program number HST-GO-15900 (PI: H. Hwang), HST-GO-16210, and HST-GO-16892 was provided by NASA through grants from the Space Telescope Science Institute, which is operated by the Association of Universities for Research in Astronomy, Incorporated, under NASA contract NAS5-26555. 
 The National Radio Astronomy Observatory is a facility of the National Science Foundation operated under cooperative agreement by Associated Universities, Inc.
 Based in part on data obtained at the W. M. Keck Observatory, which is operated as a scientific partnership among the California Institute of Technology, the University of California and the National Aeronautics and Space Administration. The Observatory was made possible by the generous financial support of the W. M. Keck Foundation. This research has made use of the Keck Observatory Archive (KOA), which is operated by the W. M. Keck Observatory and the NASA Exoplanet Science Institute (NExScI), under contract with the National Aeronautics and Space Administration. The authors wish to recognize and acknowledge the very significant cultural role and reverence that the summit of Maunakea has always had within the indigenous Hawaiian community. We are most fortunate to have the opportunity to conduct observations from this mountain.
 Based in part on observations obtained at the international Gemini Observatory (Program IDs 2020B-FT-113 and GN-2022A-Q-139; PI: X. Liu), a program of NSF’s NOIRLab, which is managed by the Association of Universities for Research in Astronomy (AURA) under a cooperative agreement with the National Science Foundation. on behalf of the Gemini Observatory partnership: the National Science Foundation (United States), National Research Council (Canada), Agencia Nacional de Investigaci\'{o}n y Desarrollo (Chile), Ministerio de Ciencia, Tecnolog\'{i}a e Innovaci\'{o}n (Argentina), Minist\'{e}rio da Ci\^{e}ncia, Tecnologia, Inova\c{c}\~{o}es e Comunica\c{c}\~{o}es (Brazil), and Korea Astronomy and Space Science Institute (Republic of Korea). This work was enabled by observations made from the Gemini North telescope, located within the Maunakea Science Reserve and adjacent to the summit of Maunakea. We are grateful for the privilege of observing the Universe from a place that is unique in both its astronomical quality and its cultural significance.
 This work makes use of SDSS-I/II and SDSS-III/IV data (http://www.sdss.org/ and http://www.sdss3.org/). 
 

 \item[Author Contributions] Y.-C.C. conducted the VLA and Keck observations, reduced and analyzed the \hst , Gemini, and VLA data, performed the GALFIT analysis and the optical and NIR spectroscopic modeling. X.L. led the study, was PI of the Chandra, Gemini, \hst , and VLA observing programs, conducted the Keck observations, and reduced the Keck data. A.F. analyzed the Chandra data and performed the {\tt BAYMAX} analysis. Y.S. was PI of the Keck observing program, conducted the Keck observations, and wrote the optical and NIR spectroscopic modeling pipeline. M.O. performed the strong lensing mass model tests. N.C. performed the cosmological simulations. X.L., Y.-C.C., A.F., and Y.S. co-wrote the manuscript with the help of N.C. and M.H.. All authors contributed to the results and commented on the manuscript.

\item[Author Information] Reprints and permissions information is available at www.nature.com/reprints. The authors declare no competing financial interests. Publisher's note: Springer Nature remains neutral with regard to jurisdictional claims in published maps and institutional affiliations. Correspondence and requests for materials should be addressed to X.L. (xinliuxl@illinois.edu).

\end{addendum}

\subsection{Code availability} 

The code used to model the {\it HST} and Keck data is publicly available at \url{https://users.obs.carnegiescience.edu/peng/work/galfit/galfit.html}. The code used for the lensing mass modeling test is publicly available at \url{https://github.com/oguri/glafic2}. The code used to perform the optical-NIR spectroscopic analysis is publicly available at \url{https://github.com/legolason/PyQSOFit}. The BAYMAX code used to model the Chandra data is available upon reasonable request.

\subsection{Data availability} 

The SDSS spectrum (Plate, 1,203; FiberID, 576; MJD, 52669) is publicly available at https://www.sdss.org/. The HST, Chandra, VLA, Gemini, and Keck data are all available through their separate public data archives (HST program numbers GO-16210 and GO-16892, Chandra GO-23700377, VLA 20B-242, Gemini 2020B-FT-113 and 2022A-Q-139). Source data are provided with this paper.

\clearpage
\noindent{\Large\bf Supplementary Information}



\subsection{Target Selection}

\obj\ is one of the dual quasar candidates selected from spectroscopically confirmed quasars in the SDSS DR7 and DR14 quasar catalogs\cite{Schneider2010,Shen2011,Paris2018} using the new astrometric technique dubbed ``Varstrometry''\cite{HwangShen2020,Shen2021}. Varstrometry utilizes the centroid jitters due to the asynchronous variability of an unresolved dual quasar. Dozens of high-redshift kpc-scale dual quasar candidates were selected by applying Varstrometry to Gaia DR2 data and follow-up \hst\ optical dual-bang imaging\cite{Shen2021,ChenHwang2022}. \obj\ is a high-piority target for detailed follow-up observations given the high redshift of $z=2.17$ and the small separation of 0\farcs46. Supplementary Table 1 lists its basic photometric and spectroscopic properties. Previous high angular-resolution VLBA observations showed two compact radio cores, confirming both sources as quasars\cite{Shen2021}. However, the alternative scenarios of a lensed quasar or a pre-merger projected quasar pair at larger physical separations could not be completely ruled out based on the previous \hst\ dual-band optical and VLBA single-band radio images.

\subsection{Comparison with Known Dual Quasar Candidates}

Supplementary Figure 1 shows \obj\ in comparison with all the previously known $1<z\lesssim3$ galactic-scale (sub-10 kpc) candidate dual quasars from the literature. Four of the candidates\cite{Anguita2018,Lemon2018,Lemon2020} are nearly identical quasars (NIQs; shown in grey) -- quasars with nearly identical optical spectra, which are generally thought to be lensed quasars lacking a definitive detection of foreground lenses due to observational limitations\cite{Lemon2020}. Three have more noticeable spectral differences\cite{Inada2008,Schechter2017,Tang2021} (shown in magenta), but without the host galaxy mergers unambiguously observed, they are still debatable as either pre-merger projected quasar pairs at larger physical separations or gravitational lenses. The dual quasar candidate\cite{Shen2021} (shown in blue) exhibits differences in the broad emission line strengths\cite{Mannucci2022} but is still controversial as a quasar lens or a pre-merger projected quasar pair at larger physical separations due to the lack of high-resolution deep NIR imaging. Finally, a $z=2.4$ quadruply lensed quasar (shown in green) has been proposed to contain a lensed kpc-scale dual quasar\cite{Lemon2022}, although the interpretation is subject to lens modeling uncertainties. Importantly, lensing can enhance angular resolution and astrometry to identify such rare systems of small-scale dual quasars that are also strongly lensed \cite{Barnacka2018}, as reported\cite{Spingola2019,Schwartz2021} in the dual AGN candidate in the lensed AGN MG B2016$+$112 at $z=3.273$ with a separation of hundreds of parsec.

\subsection{HST/WFC3 NIR Imaging}

High-resolution NIR imaging of \obj\ was conducted on 2 February 2022 UT (Program GO 16892; PI: X. Liu) with the \hst\ Wide Field Camera 3 (WFC3) in the F160W filter in a single orbit. The F160W filter operates in the wide $H$ band with a pivot wavelength of 1536.9 nm and a width of 268.3 nm\cite{Dressel2022}. The total net exposure time was 2055 s. The IR detector has a pixel size of 0\farcs13, which undersamples the PSF with a full width at half maximum (FWHM) of 0\farcs15. We dithered the observations to properly sample the PSFs while rejecting cosmic rays and bad pixels. We reduced the \hst\ data following standard procedures\cite{Sahu2021}. Geometric distortion and pixel area effects were corrected. We combined dithered frames while rejecting cosmic rays and hot pixels. The final image has a pixel scale of 0\farcs06 for the $H$-band image to be Nyquist-sampled. The photometric zeropoint is 25.936 (28.177) in the AB (ST) system, which was derived based on observations of four white dwarf standard stars\cite{Dressel2022}. Figure \ref{fig:hst_galfit} shows the \hst\ F160W image of \obj .

\subsection{Keck AO-Assisted NIR Imaging}

Keck AO-assisted NIR imaging was conducted on 21 December 2021 UT (Program N072; PI: Y. Shen) with the near-infrared camera NIRC2 in the $K_p$ filter. The $K_p$ filter has a central wavelength of 2.124 um and a bandpass width of 0.351 um. The observations were conducted in the Laser Guide Star (LGS) mode. The artificial LGS was used as a reference object to correct for atmospheric distortion. We used the star N85C000573 (located 52\farcs62 from \obj ) as the reference tip-tilt star. The images were dithered using a 4$''\times$4$''$ 9-position box. The total net exposure time was 1350 s. The individual images were reduced following standard procedures and retrieved through the Keck Observatory Archive. Individual exposures are then co-added together after correcting for the dither offsets. The final image has a pixel scale of 0\farcs01. The zeropoint of the Keck $K_p$-band image is 24.80 (Vega), which was derived based on measurements taken on AO engineering stars during laser checkouts. Supplementary Figure 2 shows the Keck image of \obj .

\subsection{Host Galaxy Morphology and Structural Decomposition}

We study host galaxy morphology and perform PSF-host structure decomposition for the \hst\ F160W (H-band) and Keck AO $K_p$ images. We adopt GALFIT\cite{Peng2010} to model the 2D surface brightness profiles for the galaxy merger. A field star is adopted to build the PSF model for the \hst\ image. For the Keck image, a standard star, S852-C, which was observed during the same night closest to \obj , is adopted to build the PSF model. Since the tip-tilt reference star is much fainter than the standard star, the strehl ratio of \obj 's Keck image is much lower. Therefore, the standard star is convolved with a wider Gaussian to match the PSF size of \obj 's observation (with a FWHM of 0\farcs05) to correct for this mismatch. 

The baseline model adopts two PSFs (for the two unresolved quasars), two S\'{e}rsic models (for the two extended host galaxies), and a constant background to fit the image. 
The S\'{e}rsic model is given by
\begin{equation}\label{eq:Sersic}
\Sigma(r) = \Sigma_e \, {\rm exp} \bigg [ - \kappa \bigg( \Big ( \frac{r}{r_e} \Big )^{1/n} -1  \bigg ) \bigg ],
\end{equation}
where $\Sigma(r)$ is the pixel surface brightness at radial distance $r$, $\Sigma_e$ is the pixel surface brightness at the effective radius $r_e$, and $\kappa$ is a parameter related to the S\'{e}rsic index $n$. $n=1$ for an exponential profile, whereas $n=4$ for a de Vaucouleurs profile. Bulge-dominated galaxies have high $n$ values ($n>2$), while disk-dominated galaxies have $n$ close to unity. We try different models until reaching the minimum reduced $\chi^2$ (i.e., $\chi_{\nu}^2$) for the smallest number of parameters. In all models considered, the centers of the S\'{e}rsic profiles were left as free parameters, but the best-fit S\'{e}rsic profile centers in all the 2 PSF + 2 S\'{e}rsic models are consistent with the best-fit PSF centers (Figure \ref{fig:hst_galfit} and Supplementary Figure 2), suggesting that the best-fit models are reasonable. The only exception was the 2 PSF + 2 S\'{e}rsic model for the Keck image, where the center of the S\'{e}rsic profile for the NE host was fixed at the NE nucleus PSF model center to help break model degeneracy given the relative shallowness of the AO imaging.

Figure \ref{fig:hst_galfit} shows the best-fit GALFIT model for the \hst\ F160W image. The extended host galaxies are detected after the PSFs are subtracted (i.e., Panel c ``Data-PSFs''). After we further subtract the extended hosts (i.e., Panel d ``Data-PSFs-Sersics''), low surface-brightness tidal features are revealed in the residual image, providing unambiguous evidence for an ongoing galaxy merger. We quantify the detection significance of the faint tidal features by using manual apertures that roughly trace the boundaries of the four tidal clumps indicated in Panel d of Figure \ref{fig:hst_galfit}. The extracted total fluxes in the aperture are all detected at $>10\sigma$ significance. Supplementary Table 7 summarizes the detection properties of the tidal features. For the brightest tidal clump T1 (SW of the pair), the detection significance is $\sim 20\sigma$, and the total AB magnitude is $25.47\pm 0.07$, which is much fainter than the host galaxies. The surface brightness depth of the HST imaging is $\sim 24.85$ AB mag/arcsec$^2$, which is consistent with the expected depth given the integration time. 

Supplementary Table 5 lists the best-fit parameters of the GALFIT models for the \hst\ F160W image. In addition to the baseline model (with 2 PSFs + 2 S\'{e}rsic), we also consider a model for the lensing scenario (Model 2; with 2 PSFs for the quasars plus 1 S\'{e}rsic for the foreground lens galaxy) and a more complicated model to attempt the bulge-disk decomposition (Model 3; with 2 PSFs for the quasars plus 4 S\'{e}rsic for the two hosts, where the S\'{e}rsic indices are fixed at $n=1$ for the disk and $n=4$ for the bulge for each host to help break model degeneracy). The baseline model is statistically preferred over the lensing model based on the $\chi^2_{\nu}$ value comparison. 

While the more complicated model (Model 3) yields a slightly smaller $\chi^2_{\nu}$ than the baseline model, the improvement is too small to warrant the larger number of parameters. For the \hst\ image, the best-fit S\'{e}rsic index in the baseline model for the host of the SW nucleus is $n=5.2$, suggesting a bulge-dominated host (but see caveats below) with an effective radius of 1.4 kpc, whereas that of the host of the NE nucleus is $n=0.9$, suggesting a disk-dominated host with an effective radius of 4.3 kpc.

Supplementary Figure 2 shows the best-fit GALFIT model for the Keck AO $K_p$ image. While the sensitivity of the Keck image is much lower than that of the \hst\ image, the Keck image offers a higher angular resolution of $\sim$0\farcs05 with much better PSF sampling (0\farcs01 pixel$^{-1}$), which is useful for better disentangling the quasars and the hosts, characterizing the host structures, and revealing any foreground lens galaxy in the lensing scenario. After we subtract the two unresolved PSFs, the extended host galaxies of the two quasars are also detected in the Keck AO image. {Importantly, most of the extended emission is centered on each of the quasar point sources, suggesting that it is associated with the host galaxies. }

Supplementary Table 6 lists the best-fit parameters of GALFIT models for the Keck AO $K_p$ image. The lensing scenario (i.e., with 2 PSFs + 1 S\'{e}rsic) is statistically ruled out based on the $\chi^2_{\nu}$ value comparison. A more complicated model is not warranted to attempt the bulge-disk decomposition for the Keck image, given its much lower sensitivity, which is not enough to detect any additional extended disks. The tidal features seen in the \hst\ image (in the GALFIT model residual) are too faint to detect in the Keck AO image.

For the Keck AO image, the best-fit S\'{e}rsic index is $n=0.2$ for the host of the SW nucleus (with an effective radius of 1.1 kpc) and $n=0.6$ for the host of the NE nucleus (with an effective radius of 1.8 kpc), suggesting disk-dominated galaxies for both hosts. The S\'{e}rsic index estimate for the NE nucleus host is broadly consistent between the \hst\ and Keck images, although the effective radius estimate is larger for the \hst\ image than that for the Keck AO image, likely due to the differences in the image resolution (i.e., the more compact disk component is better resolved with Keck AO) and sensitivity (i.e., the \hst\ image is deep enough to reveal any additional extended disk component). On the other hand, the SW nucleus host's morphological type is more controversial. The discrepancy may be caused by the different image resolutions and PSF samplings, i.e., the smaller SW host is more challenging to robustly resolve with \hst\ imaging given its lower angular resolution and poorer PSF sampling. We conclude that both hosts are likely to be disk-dominated galaxies based on the Keck-based measurements, given the higher angular resolution (PSF FWHM of 0\farcs05 for Keck compared to 0\farcs15 for \hst ) and finer PSF sampling (0\farcs01 pixel$^{-1}$ for Keck compared to 0\farcs06 pixel$^{-1}$ for \hst ) for more robustly disentangling the quasars and their host galaxies and measuring the host galaxy structures, even though the \hst\ PSF is more stable.  

Galaxy merger simulations have shown that galaxy structure plays a dominant role in regulating nuclear activity in major mergers: inflows in bulgeless galaxies occur earlier in galaxy mergers whereas stellar bulges stabilize galaxies against bar modes and inflows until the galaxies merge\cite{Mihos1996}. The disk dominance of the hosts in \obj\ may be critical for efficient SMBH fueling to produce two luminous quasars in the early-stage merger without the stabilization effect of stellar bulges which are often present in low-redshift low-luminosity dual AGNs\cite{Shangguan2016}.

\subsection{Strong Lensing Tests}

After subtracting the two quasar PSFs, we also detect significant extended emission around the two quasars in both HST and Keck AO images. At redshift $z\sim 2$, the host galaxies of luminous quasars are detectable in high-resolution near-IR imaging from HST or ground-based AO\cite{Jahnke09}. Therefore it is reasonable to attribute the extended emission to quasar host galaxies. Importantly, the Keck AO imaging has much better spatial resolution than HST to show that most of the extended emission is at the locations of the two quasars, rather than in between as resulting from a putative foreground lens galaxy. Even in the HST image (Panel c in Figure \ref{fig:hst_galfit}), the centroid of the extended emission lies closer to the brighter SW nucleus. Thus the locations of the extended emission contradict a simple lens model where the image closer to the lens is expected to be fainter. Assuming most of the detected extended emission is from the host galaxy emission, the residual map in Panel d of Figure \ref{fig:hst_galfit} shows no clear presence of a central putative lens galaxy. But the required lens galaxy with a minimum total brightness $H\lesssim 22$ (AB magnitude) would be strongly detected (see below). In addition, if the two quasars were lensed images, their host galaxy images would show noticeable distortions, which are not observed (i.e., the 2 Sersic profiles in Model 1 are well fit). The average surface brightness for the two hosts within the half-light radius is also markedly different (Supplementary Tables 5 and 6), contradictory to the strong lensing expectation. 


Alternatively, we could interpret the extended emission as mostly from a foreground lens galaxy, with limited contribution from the host galaxy of the background, strongly lensed quasar. In this case the lens galaxy would be bright, with $H\sim 20.7$ (AB magnitude). Given the image separation and source redshift, the lensing galaxy should be brighter than $H\sim 22$ AB mag irrespective of its redshift as predicted from the Faber-Jackson relation\cite{Rusin2003}. A lens galaxy as bright as $H\sim 20.7$ (AB) would be possible if the lens galaxy were at redshift $z\sim 0.3$ or $z\sim 1.8$ (the two solutions from the lens model). This way we could reconcile the lack of distorted morphology of the lensed host galaxy and the difference in the surface brightness of the lensed images. However, the locations of the extended emission, and the expectation of detecting significant host emission in $z\sim 2$ quasars are at odds with this alternative scenario. For example, in this scenario, the estimated total stellar mass of the quasar host would be much lower, resulting in exceedingly massive black holes with respect to the host stellar mass.

To more quantitatively test the strong lensing hypothesis where most of the extended emission comes from the foreground lens galaxy, we perform strong lensing mass modeling tests using the glafic software\cite{Oguri2010} for both the {\it HST} and the Keck images where the putative lens galaxy's position and mass can both be constrained with GALFIT. In this scenario, what we observe in the \hst\ image would be two quasar images (two PSFs) plus the lens galaxy (extended emission in the PSF-subtracted residuals). We then assume that ``Model 2'' listed in Supplementary Table 5 (from the GALFIT analysis of the \hst\ image) tells us the quasar image positions as well as the position of the lens galaxy in the lensing scenario. We have seven constraints from the observation, i.e., the positions of the two quasar images (4), the flux ratio (1), and the position of the lens galaxy (2). The strong lensing mass modeling is illustrated in Supplementary Figure~3.

Our baseline model assumes a simple singular isothermal ellipsoid (SIE) lens. The SIE model contains seven parameters, i.e., the mass strength (1), the lens position (2), the ellipticity and its position angle (2), and the source position (2). Therefore, the adoption of an SIE model is justified given the number of constraints.

We assume positional errors of 0\farcs01 for PSFs and 0\farcs03 for the lensing galaxy. These positional uncertainties were inflated by a factor of $\sim 5$ from the GALFIT uncertainties, which are statistical only and likely underestimated. Adopting a positional error of 0\farcs05 instead for the lensing galaxy does not affect the conclusion but makes the fitting more unstable. Assuming an SIE lens model, we were able to obtain a mass model that fit the positions and flux ratio of the double optical nuclei, so the lensing hypothesis cannot be rejected based on the Model 2 best-fit positions for the \hst\ image (Supplementary Table 5) alone. However, Model 1 is statistically favored over Model 2 based on GALFIT results, which means that the dual quasar scenario is preferred by the data over the strong lensed quasar scenario, even though the quantitative strong lensing mass modeling test cannot rule out lensing (where the extended emission is assumed to be the foreground lens). In addition, the residual map of Model 2 (the strong lensing hypothesis) shows excess fluxes at the locations of the quasars, indicating host galaxy emission that is not properly accounted for by the model. 

A more definitive test of the strong lensing hypothesis comes from the Keck AO imaging, given its higher resolution and better localized extended emission. We performed the same lensing mass modeling for the Keck image using glafic, assuming positional errors of 0\farcs01 for PSFs and 0\farcs03 for the lensing galaxy. Assuming a simple SIE lens model, we were unable to obtain a mass model that could fit the positions and flux ratio of the double optical nuclei. With a more complicated SIE plus shear model, we got a reasonably good fit, but it would require very large external shear (with a best-fit value of $\sim$0.6) and therefore quite unrealistic. Thus our mass modeling test suggests that the Model 2 best-fit positions for the Keck AO image (Supplementary Table 6) are inconsistent with the lensing hypothesis. 

Supplementary Table 8 lists all the parameters (mean values and 1$\sigma$ uncertainties) obtained using glafic for the SIE or the SIE+shear models from the Markov Chain Monte Carlo (MCMC) analysis for both the \hst\ and Keck images. Since the number of degree of freedom for the SIE model is zero, if the model is reasonable, we expect to have $\chi^2_{{\rm min}}{\sim}$0 for the best-fit model. If this is not the case, it means that either the assumption on the lens mass model is incorrect, or it is not a strong gravitational lens system. For some of the parameters, the posterior distribution is not necessarily Gaussian, so the mean can be different from the best-fit value (e.g., in the Keck SIE+shear model, the best-fit value of external shear is $\sim$0.6 but the mean is 0.48$\pm$0.12).

\subsection{HST/STIS Optical Slit Spectroscopy}

Spatially resolved optical slit spectroscopy was conducted for \obj\ on 16 February 2021 UT (Program GO 16210; PI: X. Liu) with the \hst /Space Telescope Imaging Spectrograph (STIS) using the G750L grating. A slit width of 0\farcs2 was used. The slit direction was oriented at P.A. = 236.4$^{\circ}$ (E of N) to simultaneously cover the two nuclei. The total net exposure time was 2136 s. The data were reduced using the $stis\_cti$ package, which removes trails and other artifacts caused by Charge Transfer Inefficiency effects\cite{Anderson2010}. The exposures were flat-fielded and combined to reject cosmic rays. The final 1-dimensional (1D) spectra for each nucleus were extracted using an aperture size of 7 pixels, corresponding to 0\farcs35. Supplementary Figure 4 shows the {\it HST}/STIS 2-dimensional (2D) spectrum and demonstrates the extraction of the 1D spectra for each nucleus. The two nuclei are well separated by HST/STIS. Figure \ref{fig:OIR_spec} shows the final STIS spectra for the double nucleus.

\subsection{Gemini/GMOS Optical Slit Spectroscopy}

Gemini spatially resolved optical slit spectroscopy was conducted on April 20 2022 UT (Program GN-2022A-Q-139; PI: X. Liu) with the Multi-Object Spectrographs (GMOS) on Gemini-North. The R150 grating and a slit width of 0\farcs5 were used, yielding a spectral resolution of $R\sim631$. The slit direction was oriented at P.A.=236.1$^{\circ}$ (E of N) to simultaneously cover the two nuclei. The total net exposure time was 1876 s. The average seeing during the observations was $\sim$0\farcs49 in the optical wavelengths. The spectra cover the wavelength range of 400-1200 nm. The data were reduced using the Gemini PyRAF/IRAF pipeline following standard procedures. The reduction steps included bias subtraction, flat-field correction, cosmic-ray rejection, wavelength calibration, and flux calibration. The CuAr spectra with the same instrumental configurations were used for the wavelength calibration. EG131 was observed as the standard star for spectrophotometric calibration.

Supplementary Figure 5 shows the Gemini/GMOS 2D spectrum and demonstrates the source decomposition of the 1D spectra for each nucleus. Based on the 1D spatial profile of the wavelength-averaged spectrum, the two nuclei are well modeled by two Gaussian components whose separation is consistent with that observed in the {\it HST} optical images. We have decomposed the two nuclei using two Gaussians using a fixed separation of 0\farcs46 across all wavelengths to help with the decomposition. The FWHM of the Gaussian components is a free parameter during the fitting process to account for the variable seeing as a function of wavelength. Figure \ref{fig:OIR_spec} shows the final decomposed GMOS spectra for the double nucleus.

\subsection{Gemini/GNIRS NIR Slit Spectroscopy}

Gemini spatially resolved NIR slit spectroscopy was conducted on 22 December 2020 UT (Program GN-2020B-FT-113; PI: X. Liu) with the Near Infra-Red Spectrograph (GNIRS) on Gemini-North. We used the cross-dispersion (XD) 32/I mode with the blue camera and a slit width of 0\farcs45, yielding a spectral solution of $R\sim1000$. The total net exposure time was 2400 s. The average seeing during the observations was $\sim$0\farcs34 in the NIR wavelengths. The detector controller artifacts in the raw images were removed using the {\tt cleanir.py} script. Then, the data were reduced with the Gemini IRAF pipeline following standard procedures. The reduction steps included flat-field correction, S-distortion correction, wavelength calibration, telluric correction, and flux calibration. The pinhole images were used to correct the S-distortion in the images and then the Argon spectra were used for the wavelength calibration. HIP26000 was observed as the standard star for telluric correction and spectrophotometric calibration. 

Supplementary Figure 6 shows the Gemini/GNIRS 2D spectrum and demonstrates the source decomposition of the 1D spectra for each nucleus. Based on the 1D spatial profile of the wavelength-averaged spectrum, the two nuclei are marginally resolved. The 1D profile is consistent with two Gaussian components with separation of 0\farcs46, although a third faint extended component is seen near the SW nucleus, which is likely due to the extended host galaxy emission in the merger. We adopted the local minimum between two Gaussian as the central aperture boundaries. The final 1D spectra for each nucleus were extracted using a boxcar aperture with a size of 21 pixels (corresponding to 1\farcs05) for the SW nucleus and 7 pixels (corresponding to 0\farcs35) for the NE nucleus. The aperture for the NE nucleus does not fully cover the tail of the emission ($>$91 pixels) because it reaches the edge of the slit. Figure \ref{fig:OIR_spec} shows the final decomposed GNIRS NIR spectra for the double nucleus.

\subsection{Joint Spectral Modeling of the HST/STIS and Gemini/GNIRS Data}

We jointly fit the optical and NIR spectra of each nucleus to estimate the systemic redshift and measure emission-line properties for each quasar. We use the PyQSOFit fitting code\cite{Shen2021} following the procedures as described in detail in\cite{Shen2012}. Here we briefly summarize the fitting procedures and model assumptions. Each combined spectrum has been de-reddened for Galactic extinction using the Milky Way reddening law\cite{cardelli89} and $E(B-V)$ derived from the dust map\cite{Schlegel98}. The spectrum is modeled as a linear combination of a pseudo-continuum (consisting of a power-law plus polynomial quasar continuum and Fe {\small II} emission), broad, and narrow emission lines. A Balmer continuum component has also been considered in the models but made negligible contribution in the best-fit results. First, to estimate the systemic redshift, we perform a global fitting to the joint optical-NIR spectrum. Given the moderate S/N of the spectra, the velocities of all the narrow and broad emission lines have been tied to be the same to help break model degeneracy. The best fit systemic redshift is $2.1679\pm0.0003$ for the SW nucleus and $2.1665\pm0.0010$ for the NE nucleus which we take as our baseline estimates. Alternatively, allowing for a nonzero velocity offset between the narrow and broad emission lines yields a best fit systemic redshift of $2.1686\pm0.0007$ for the SW nucleus and $2.1653\pm0.0021$ for the NE nucleus based on the narrow emission lines, which are consistent with the baseline estimates within uncertainties. The redshift uncertainties are statistical uncertainties only as quasar emission lines tend to have different velocity shifts from the systemic velocity that depend on the line species\cite{Shen_etal_2016b}. Nevertheless, these statistical uncertainties are still useful to assess the redshift difference for the strongly lensed quasar scenario.

Second, to make emission-line measurements and derive physical quasar properties, we perform local fittings for each emission-line complexes. Given the moderate S/N of the existing spectra, each narrow and broad emission line component is modeled by a single Gaussian, where an upper limit of FWHM = 1200 km s$^{-1}$ has been imposed on the narrow lines. The velocities of all the narrow emission lines have been fixed to be the same as the best-fit systemic redshift of each quasar, whereas non-zero velocity offsets are allowed for each broad emission lines (i.e., with respect to the systemic redshift) to better account for possible velocity offsets in different broad lines due to ionization stratification, accretion disk winds, and/or IGM absorption. The widths of all the narrow emission lines are tied to be the same to help break model degeneracy, but the widths of the broad emission lines are allowed to vary among different lines. The flux ratios of the \OIII\ and \NII\ doublets are fixed to be their theoretical values to help break model degeneracy and reduce ambiguities in decomposing the emission line complexes. Supplementary Figure 7 shows the best-fit spectral models for both the global and the local fits. Supplementary Table 2 lists the emission-line measurements from the local best-fit models.

\subsection{Chandra ACIS-S X-ray 0.5--8 keV Imaging Spectroscopy, X-ray Data Reduction, and Data Analysis Using BAYMAX}

We observed \obj\ with the ACIS-S on board the \chandra\ X-ray Observatory on 23 Dec 2021 UT (Program 23700377, ObsID=25710 and 26245; PI: X. Liu). \obj\ was observed on axis on the S3 chip with a total exposure time of 30 ks. We follow a standard data reduction procedure for each observation using \emph{Chandra} Interactive Analysis of Observations software ({\tt CIAO}) v4.13\cite{Fruscione2006}. We correct for astrometry by cross-matching \emph{Chandra}-detected point-like sources with available optical catalogs. We require a minimum of 3 matches between the X-ray observations and optical catalogs, and we required each matched pair to be separated by less than 2$''$. For each \chandra\ observation, we find background flaring to be negligible, with no time interval containing a background rate 3$\sigma$ above the mean level. We restrict our analysis to photons with energies between 0.5--8 keV. We analyze the photons that are contained within a 20$''\times$20$''$ box centered on the nominal X-ray coordinates of \obj . Supplementary Figure 8 shows the binned \chandra~data, with the soft X-rays (0.5--2 keV) shown in red on the left (total of 114 counts) and the hard X-rays (2--8 keV) shown in purple on the right (total of 126 counts). 

We analyze the \chandra\ observations for the presence of two X-ray point sources using {\tt python} tool {\tt BAYMAX} (Bayesian AnalYsis of Multiple AGN in X-rays)\cite{Foord2019,Foord2020,Foord2021a}. {\tt BAYMAX} allows for a quantitative analysis of whether a source in a given \emph{Chandra} observation is more likely composed of one or two point sources. The main component of {\tt BAYMAX} is the calculation of the Bayes factor $\mathcal{B}$, which represents the ratio of the plausibility of observed data, given each model. {\tt BAYMAX} takes calibrated \emph{Chandra} events and compares them to the expected distribution of counts for a single point source ($M_{1}$) versus a dual point source ($M_{2}$) model. 

When calculating the Bayes Factor, {\tt BAYMAX} uses {\tt python} nested sampling package {\tt nestle}, where both probability and prior densities need to be defined. The probability densities of $M_{1}$ and $M_{2}$ are estimated by comparing the \chandra\ sky coordinates ($x$, $y$) and energies ($E$) of each detected X-ray event to simulations based on single and multiple point source models\cite{Foord2019}. The prior vectors for $M_{1}$ and $M_{2}$ are $\theta_{1} = \{\mu_{1}, f_{{\rm bkg}}\}$ and $\theta_{2} = \{\mu_{1}, \mu_{2}, f, f_{{\rm bkg}}\}$, respectively. Here, the sky-x, sky-y coordinates of each point source (encompassed by $\mu$) are uniform distributions across the entire 20$''\times$20$''$ field of view. The parameter $f_{{\rm bkg}}$ represents the number of counts associated with a spatially uniform background component versus a point-source component; the prior density is represented in log-space with log-normal distribution with a mean of $-2$ and a large standard deviation ($\sigma$=0.5) to allow flexible movement in prior space. Lastly, the parameter $f$ represents the count ratio between the secondary and primary point-source; the prior density if represented in log-space with a log-normal distribution set between $-4$ and $0$. For sources with more than one \chandra\ observations, as in the case of \obj , the code also fits for astrometric shifts between each observation, although the point spread function model used in the probability densities are unique to each observation.

These are the standard ``non-informative" priors used by {\tt BAYMAX} when analyzing X-ray point sources\cite{Foord2020,Foord2021a}. Running {\tt BAYMAX} on the \chandra\ observations of \obj\ we find log $\mathcal{B} = 3.3 \pm 0.7$. The error bars are returned by {\tt nestle} and represent the statistical sampling error of the nested sampling procedure. Previous analyses have found this error to be consistent with the $\sim68\%$ confidence interval when running {\tt BAYMAX} 100 times on the same source\cite{Foord2020}.

In addition to calculating a Bayes factor, {\tt BAYMAX} estimates optimal values for parameters such as the separation ($r$) and count ratio ($f$) using {\tt python} MCMC package {\tt PyMC3}. We find best-fit values of $r=1\farcs54_{-0\farcs36}^{+0\farcs39}$ and log$f=-1.43_{-0.30}^{+0.30}$, at the 95\% confidence interval. The overestimation of $r$ as compared to the resolved separation of the two optical cores (along with the large error-bars) are likely a result of the secondary X-ray point source having, on average, only 6 counts. The low number of counts associated with a putative secondary X-ray point source ($\sim$6) affects the accuracy of estimations on $r$. However, the estimated position angle between the two X-ray point sources ($350_{-24}^{+37}$ deg.; 95\% confidence range) is consistent with the two resolved cores in the \emph{HST} imaging (330 deg.). Deeper \emph{Chandra} observations can help better discern whether the X-ray emission is spatially consistent with the \emph{HST} imaging.

Sampling from the posterior distributions, the code will probabilistically assign counts to different model components (which, for $M_{2}$ is the primary point source, the secondary point source, and a background component), from which a spectra can be created and analyzed. This allows for the individual spectral fitting of each X-ray point source in a putative AGN pair. Iterating this process results in a set of 100 spectral realizations for each model component. We fit and model each spectral realization, for each X-ray point source, using {\tt XSPEC}\cite{Arnaud1996}. This allows us to quantify the average values of the spectral shape and X-ray luminosity, across all spectral fits. With the fitted X-ray spectra, we estimate the hard X-ray (2--8\,keV) flux ratio of the two nuclei (NE/SW) to be $45^{+41}_{-33}$.

Supplementary Figure 9 shows the spectral fits to the primary (left) and secondary (right) X-ray point source detected by {\tt BAYMAX}.  We find both sets of spectral realizations are best-fit with a simple absorbed power law ({\tt XSPEC} model components: {\tt phabs} $\times$ {\tt zphabs} $\times$ {\tt zpow}). Given the low number of counts for both components (average of $\sim191.5$ and $\sim6$ counts for the primary and secondary, respectively), we fix the power law index, $\Gamma$, to a value of 1.8 for all spectral fits. Supplementary Table 3 lists the median values and the confidence interval of the best-fit spectral parameter distributions.

\subsection{X-ray-to-optical Luminosity Ratio as Characterized by the Spectral Slope $\alpha_{{\rm OX}}$}

The slope of a hypothetical power law from 2500 \AA\ to 2 keV is usually defined by $\alpha_{{\rm OX}}=0.3838 \log (l_{2500\,\AA}/l_{2\,keV})$, where $l_{2500\,\AA}$ is the rest-frame monochromatic optical luminosity (in units of erg s$^{-1}$ Hz$^{-1}$) at 2500 \AA\ measured from the dereddened spectrum for each nucleus and $l_{2\,keV}$ is the rest-frame 2 keV luminosity density (in units of erg s$^{-1}$ Hz$^{-1}$) \cite{Green2009}. The X-ray-to-optical spectral indices $\alpha_{{\rm OX}}$ are calculated as 1.46$\pm$0.01 for the NE nucleus and 2.32$\pm$0.13 for the SW nucleus. As an intrinsic probe of the accretion processes, $\alpha_{{\rm OX}}$ characterizes the relation between the optical/UV thermal blackbody radiation from the accretion disk and the X-ray emission from the hot corona. $\alpha_{{\rm OX}}$ of optically selected quasars has been observed to strongly correlate with optical luminosity with a possible residual evolution with redshift\cite{Green2009}. Compared with the typical optically selected quasars at similar luminosities\cite{Green2009}, $\alpha_{{\rm OX}}$ of the NE nucleus is consistent the general population. On the other hand, $\alpha_{{\rm OX}}$ of the SW nucleus (which also has a higher estimated Eddington ratio) is much larger, which may be evidence for the SMBH experiencing some unusual accretion state induced by the galaxy merger. Alternatively, the large $\alpha_{{\rm OX}}$ of the SW nucleus could be evidence of high circumnuclear absorption as expected in galaxy mergers, although the absorption material would have to be at the smallest scales that only affect the X-ray emitting region but do not obscure the optical emitting region given the Type 1 nature of the optical nucleus. 

\subsection{VLA Observations, Data Reduction, and Measurements}

Dual-band radio imaging was conducted on 11 and 12 December 2020 UT (Program 20B-242; PI: X. Liu) with the VLA in its most extended A configuration. C-band (6.0 GHz) and Ku-band (15.0 GHz) were adopted with an on-source time of 0.5 hours for each band. The images were calibrated and reduced through the standard VLA Calibration Pipeline version 5.6.2 in the Common Astronomy Software Applications package (CASA)\cite{McMullin2007}. Sources 3C147 and 3C286 were used as the flux and bandpass calibrators. Source J0738+1742 was used as the gain calibrator as well as the phase reference source in the Ku-band. We fit 2D Gaussian components to each nucleus to obtain the flux densities. Two nuclei are well resolved in the Ku-band image and are consistent with two point sources. The two nuclei are marginally resolved in the C-band image, therefore, we fit the C-band image using two Gaussian components with a fixed shape, same as the synthesized beam, to obtain accurate flux estimations. Figure \ref{fig:vla} shows the VLA A-config two-band radio images for \obj . Supplementary Table 4 lists the radio measurements.

\subsection{Estimation of Virial Black Hole Masses, Bolometric Luminosities, and Eddington Ratios}

We estimate SMBH properties for each nucleus using the combined GMOS+GNIRS spectra. These include SMBH masses, bolometric luminosities, and Eddington ratios. The GMOS spectrum is adopted in our baseline estimates because of the higher S/N than that of the STIS spectrum. 

We estimate SMBH masses using the single-epoch virial mass estimator\cite{Shen2012}. The virial SMBH mass is estimated as 
\begin{equation}
    \text{log}_{10} \bigg(\frac{M_{\text{BH}}}{M_{\odot}}\bigg) = a + b \text{ log}_{10}\bigg(\frac{\lambda L_{\lambda}}{10^{44}\,\text{erg s}^{-1}}\bigg) + c \text{ log}_{10}\bigg(\frac{\text{FWHM}}{\text{km s}^{-1}}\bigg),
\end{equation}
where $L_\lambda$ is the monochromatic continuum luminosity at wavelength $\lambda=$5100\AA, FWHM is the full width at half maximum of the \halpha\ line, and $a=1.390$, $b=0.555$, $c=1.873$ are the coefficients from the calibration of \cite{Shen2012}. The \halpha-based virial BH masses are ${\sim}10^{9.1\pm0.4}M_{\odot}$ for the NE nucleus and ${\sim}10^{9.2\pm0.4}M_{\odot}$ for the SW nucleus, whereas the errors indicate total 1$\sigma$ uncertainties including systematic errors. We have also estimated the virial BH masses using the broad \MgII\ lines as an independent double check. The \MgII-based BH masses are estimated as ${\sim}10^{8.9\pm0.4}M_{\odot}$ for the NE nucleus and ${\sim}10^{9.3\pm0.4}M_{\odot}$ for the SW nucleus, which are consistent with the \halpha-based BH masses within the systematic uncertainty of $\sim$0.4 dex\cite{Shen2013}. 

We estimate the bolometric luminosity using the rest-frame optical flux density at 5100 \AA\ and the relevant Bolometric Correction (BC)\cite{Richards2006}. The bolometric luminosity is estimated as 
\begin{equation}
    L_{\rm bol} = \lambda L_{\lambda} \times BC,
\end{equation}
where $L_\lambda$ is the monochromatic continuum luminosity at rest-frame wavelength $\lambda$=5100 \AA\ and BC = 10.33 is the bolometric correction at 5100 \AA \cite{Richards2006}. The bolometric luminosities are estimated as $\sim$(2.0$\pm$0.4)$\times$10$^{46}$ erg/s for the NE nucleus and $\sim$(5.0$\pm$1.0)$\times$10$^{46}$ erg/s for the SW nucleus.

To infer the Eddington ratio, we calculate the Eddington luminosity from the BH mass as,
\begin{equation}
    L_{\rm Edd} \sim 1.26 \times 10^{38} \bigg(\frac{M_{\rm BH}}{M_{\odot}}\bigg)\ { \rm erg s}^{-1}.
\end{equation}
 The Eddington ratios $L_{bol}/L_{Edd}$ are estimated as $\sim$0.12$\pm$0.03 for the NE nucleus and $\sim$0.25$\pm$0.07 for the SW nucleus.

In addition to the inferences based on the optical measurements, we also estimate the bolometric luminosities and Eddington ratios based on the X-ray measurements as an independent double check. We use the X-ray luminosity $L_{2-8\,{\rm keV}}$ and apply the X-ray bolometric correction for radio-loud quasars (Equation 14 in \cite{Runnoe2012}) . The  estimated bolometric luminosities are $\sim$10$^{46.59^{+0.03}_{-0.02}}$ erg/s for the NE nucleus and $\sim$10$^{45.72^{+0.17}_{-0.30}}$ erg/s for the SW nucleus. The inferred Eddington ratios are $\sim$0.23$^{+0.04}_{-0.03}$ for the NE nucleus and  $\sim$0.03$^{+0.01}_{-0.02}$ for the SW nucleus. The bolometric luminosities and Eddington ratios are in broad agreement with the optically based estimates considering substantial scatters in the X-ray bolometric correction\cite{Runnoe2012}.

\subsection{X-ray-to-Radio Multi-wavelength Spectral Energy Distributions}

Supplementary Figure 10 shows the multi-wavelength spectral energy distributions (SEDs) for both nuclei in \obj . The data include photometry in the radio, infrared, optical and X-ray wavelengths, as well as the optical and NIR spectra. For context, also shown are various optically-selected quasar SED templates from the literature\cite{Richards2006,Shang2011}. Overall, the SEDs of both nuclei are consistent with the SED templates of typical optically selected (single) quasars. Both nuclei are classified as radio-loud quasars.

\subsection{Estimation of Host-galaxy Stellar Masses} 

We estimate the total stellar mass for the host galaxy of each nucleus using the \hst\ F160W and Keck $K_p$-band PSF-subtracted host-only magnitudes. We first obtain the \hst\ F160W magnitudes and Keck $K_p$-band magnitudes of the host galaxies by decomposing the 2D surface brightness profiles using GALFIT. We then run the stellar population synthesis code GALAXEV\cite{BC03} to create composite stellar population (CSP) models to fit the observed two-band host-only photometry. While only two-band data are available for the hosts, they allow a reasonable estimate of the mass-to-light ratio (M/L) to estimate the stellar mass. The BaSeL spectral library, a Chabrier initial mass function (IMF)\cite{Chabrier2003}, and an exponentially decaying star formation history (SFH) are assumed to build the CSP SED. The evolution tracks for three different metallicities ($Z$=0.02, 0.008, and 0.004)\cite{Fagotto1994,Bressan1993} are adopted to model any CSP change due to different metallicities. Dust attenuation assuming a fixed optical depth is adopted. We create a grid of CSPs from assuming different stellar population parameters. The fitted free parameters include the total stellar mass, age, metallicity, the exponential decay star formation timescale, and the dust optical depth. The CSP models are fit to the PSF-subtracted \hst\ F160W and Keck $K_p$-band magnitudes to estimate the best-fit stellar population parameters of the host galaxies. 

Supplementary Figure 11 shows the best-fit CSP models and the PSF-subtracted host-only two-band photometry. The total stellar masses are estimated as $\sim$10$^{11.46^{+0.02}_{-0.26}}$ $M_{\odot}$ for the NE nucleus and 10$^{11.50^{+0.04}_{-0.27}}$ $M_{\odot}$ for the SW nucleus, whereas the errors include 1$\sigma$ statistical uncertainties and the partial systematics from surface brightness profile fitting. This partial systematic uncertainty of total stellar masses is estimated to be $\sim$0.2 dex from assuming different models during the surface brightness decomposition (e.g., 2 S\'{e}rsic versus 1 S\'{e}rsic for each host) and should be taken as a lower limit since it does not include the additional systematics from adopting different assumptions about the IMF, the SFH, or the unknown dust configuration, which may result in an additional ${\sim}0.3$ dex uncertainty at least\cite{Conroy2013}.

\subsection{Black Hole Mass versus Stellar Mass Scaling Relations} 

To put \obj\ into context, we study the relation between BH mass and host-galaxy total stellar mass by comparing \obj\ with observations of the host galaxies of single quasars and AGNs, as well as inactive galaxies in the literature. Supplementary Figure 12 shows the relation between BH mass and host total stellar mass for \obj , in comparison with various literature samples. Included for comparison are the local samples of both inactive and active galaxies\cite{KormendyHo2013,Bennert21,Reines15} and samples of AGNs and quasars at $z>0.2$ selected with various methods\cite{Jahnke09,Dong16,Mechtley16,Suh20,Li21}. 
Overall, the two nuclei of \obj\ are broadly consistent with both the local scaling relations for inactive galaxies\cite{KormendyHo2013} and \hst -imaged AGNs\cite{Bennert21} and the high-redshift quasar samples, considering the large systematic errors of our measurements. The hosts of \obj\ already follow the local scaling relations without prominent stellar bulges. This suggests that at least some SMBHs may have formed before their host stellar bulges, in contrast to naive expectations from the canonical SMBH-bulge co-evolution scenario where SMBHs and their host galaxy stellar bulges form coevally from merger-induced starbursts and quasar fueling\cite{DiMatteo2005,DiMatteo2008}.

Given the high redshift and high luminosities of \obj , it is challenging to construct a large tailored sample of control quasars for a fairer comparison with similar selection methods, properties, and data qualities. The high angular resolution images in the rest-frame optical are expensive to obtain for even larger samples. Previous works  have studied optical selected quasars and estimated the total stellar mass from decomposing \hst\ NIR images\cite{Mechtley16,Li21}. However, some of the studied quasars\cite{Li21} are at significantly lower redshifts ($0.2<z<0.6$) than that of \obj . Others are focused on quasars with high SMBH masses (10$^9-$10$^{10}M_{\odot}$) only\cite{Mechtley16}. While some work also studied optical selected quasars\cite{Dong16}, the stellar mass was estimated from fitting the broad-band SED rather than spatially decomposing the PSF and the host. Some works were based on X-ray selected AGNs in the COSMOS field\cite{Jahnke09,Suh20}, which are likely to trace a different SMBH population than optically selected quasars given that X-ray selected AGNs in small deep extragalactic fields tend to have lower bolometric luminosities than typical optically selected bright quasars from wide-area surveys.

\subsection{Estimation of Merging Timescales} 

Chandrasekhar dynamical friction timescale arguments are adopted to estimate characteristic timescales for the galaxy-galaxy merger and the inspiraling timescales of the two SMBHs. Assuming singular isothermal spheres for the host galaxy density profiles and circular orbits, the dynamical friction timescale for a satellite galaxy with velocity dispersion $\sigma_{{\rm S}}$ inspiraling from radius $r$ in a host galaxy with velocity dispersion $\sigma_{{\rm H}}$ is estimated as\cite{Chandrasekhar1943}
\begin{equation}\label{eq:t_galfric}
t^{{\rm gal}}_{{\rm fric}} = \frac{2.7}{\ln \Lambda}\frac{r}{30\,{\rm kpc}}\bigg(\frac{\sigma_{{\rm H}}}{200\, {\rm km\,s^{-1}}}\bigg)^2\bigg(\frac{100\, {\rm km\,s^{-1}}}{\sigma_{{\rm S}}}\bigg)^3 ~{\rm Gyr},
\end{equation}
where $\Lambda$ is the Coulomb logarithm with typical values of 3${\lesssim}\ln \Lambda{\lesssim}$30\cite{binney87}. We assume $\ln \Lambda{\sim}$2, which is appropriate for equal mass mergers\cite{dubinski99}.  Assuming the SW galaxy as the host (primary) and the NE one as the satellite (secondary), the two galaxies would merge in $t_{{\rm fric}}{\sim}$7$^{+64}_{-5}$ Myr, corrected for projection effects assuming random orientation (i.e., $r=r_p/\sin{<I>}$, where $r_p=3.8$ kpc is the projected separation between the two galaxies and the median inclination angle is $<I>=60^{\circ}$ for random orientations). 

After merging together with their individual hosts, the two SMBHs would inspiral to the center of the merged galaxy under dynamical friction with the stellar background. Assuming a singular isothermal sphere for the density distribution of the merged host galaxy, the dynamical friction timescale of a BH of mass $M_{\rm BH}$ on a circular orbit of radius $r$ is estimated as
\begin{equation}\label{eq:t_bhfric}
t^{{\rm BH}}_{{\rm fric}} = \frac{19}{\ln \Lambda}\bigg(\frac{r}{5\,{\rm kpc}}\bigg)^2\frac{\sigma_{{\rm H}}}{200\,{\rm km\,s^{-1}}}\frac{10^8\,M_{\odot}}{M_{\rm BH}}~{\rm Gyr}, 
\end{equation}
where $\ln \Lambda{\sim}$6 for typical values\cite{binney87}. We estimate $t_{{\rm fric}}{\sim}$0.22$^{+0.50}_{-0.16}$ Gyr for the SMBH in the NE nucleus reaching the center of the merger products and forming a gravitationally bound binary SMBH system with the SW nucleus, assuming a radius of $r=4$ kpc at the start of the orbital decay. 

The effects of friction by gas have been neglected in the estimation, which will accelerate the merger processes. On the other hand, the effects of tidal stripping of stars in the satellites have been neglected, which will delay the merger. The effect of BH accretion have also been neglected in the simple estimation, which may change the masses of the BHs and the resulting merger timescales. The actual merger involving the stellar and BH components with gas friction and gas accretion onto two SMBHs is likely much more complicated. Numerical simulations with model parameters tailored to those of \obj\ are needed to make more realistic predictions of the subsequent merger of the system\cite{renaud10}. Regardless of the actual sequence of the mergers, the SMBHs will form a gravitationally interacting bound binary system.

\subsection{Implications on the Abundance of $z\sim2$ Kpc-Scale Dual Quasars}

Observational constraints on the occurrence rate of kpc-scale dual quasars at cosmic noon are still accumulating\cite{Silverman2020,Shen2022}. A recent study using Gaia DR3 resolved pairs around luminous quasars at $z\sim2$ reported a dual quasar fraction of $\sim 6\times 10^{-4}$ over $\sim 3-30\,{\rm kpc}$ pair separations\cite{Shen2022}. About one third of the sample have pair separations less than $\sim 10\,{\rm kpc}$. In this measurement both members of the pair are optically-unobscured quasars with similar physical properties as SDSS J0749+2255. Some of the double quasars in the sample\cite{Shen2022} are gravitationally lensed quasars, which require further confirmation in follow-up observations; therefore the quoted dual quasar fraction is an upper limit. Confirming \obj\ as a genuine kpc-scale dual quasar is important in the sense that it implies at least some of the sub-arcsec double quasars at $z\sim 2$ are physical pairs, thus justifying dedicated follow-up observations to compile a statistical sample of such dual quasars at cosmic noon.

\subsection{Comparison with Cosmological Hydrodynamic Simulations}

To compare \obj\ with theoretical predictions for similar systems, we utilize the very recent large-volume, high-resolution (gravitational softening of $1.5\,{\rm kpc}/h$; dark matter mass resolution of $9.6\times 10^6\,M_\odot$) cosmological hydrodynamic simulation \texttt{ASTRID}\cite{Bird2022,Ni2022,ChenN2022astrid}.
\texttt{ASTRID} includes subgrid models for star formation, SMBH accretion, and the associated supernova and AGN feedback, recently updated with an SMBH seed population between $3\times 10^4\,M_\odot/h$ and $3\times 10^5\,M_\odot/h$ and a sub-grid dynamical friction model to follow the SMBH dynamics down to kpc scales.
With a comoving volume of $(250\,{\rm Mpc}/h)^3$, \texttt{ASTRID} is the largest galaxy formation simulation up to date that covers the epoch of the cosmic noon, and can provide a statistical sample of the rare quasar pair population down to galactic scales.

By searching through a redshift range of $z=2.0$--$2.3$ in $\texttt{ASTRID}$, we have identified 8 quasar pairs with $L_{\rm bol}>10^{46}~{\rm erg/s}$ for both quasars in the pair and with 3D spatial separations of $\Delta r < 30\, {\rm kpc}$.
Out of these 8 close quasar pairs, 6 only become both luminous ($L_{\rm bol}>10^{46}~{\rm erg/s}$) after the host galaxy merger and when the pair separation is below $\sim 2\,{\rm kpc}$.
Only two pairs are in the early stages of the galaxy merger and hence have separate host galaxies, similar to \obj .
For both pairs (PairA and PairB) we find that the quasars each contain SMBHs with masses of $M_{{\rm BH1,2}} \sim [0.8-4] \times 10^9 \,M_\odot$, and are embedded in galaxies with $M_{\rm gal} \sim$ a few $ 10^{11}\,M_\odot$.

Supplementary Figure 13 shows the $M_{\rm BH}-M_{\rm gal}$ relation for the eight bright quasar pairs. The colored lines mark PairA and PairB with apparently distinct galaxy bulges, although the kinematic bulge-disk classification adopted by simulations cannot be reliably used in interacting systems, whose morphology is typically disturbed and the disks are in the process of being disrupted in the merger process. PairB is the closest match with \obj\ in both the SMBH masses and the galaxy stellar masses.

In order for the pairs to be the closest match with \obj\ (i.e. $L_{\rm bol1,2}>10^{46}~{\rm erg/s}$ and in separate galaxies), they are likely in an early phase of the galaxy merger, and in particular at their first few pericentric passages.
This is a stage where both the quasar luminosities and the host galaxy star formation are enhanced from the merger-induced gas inflow.
Supplementary Figure 14 shows the orbits of the two quasars in one of such pairs along with the host galaxy merger.
This pair has a very eccentric orbit with an eccentricity of $0.94$ before the merger, and significant host galaxy tidal disruption is seen, after which the secondary quasar becomes offset from the galaxy merger remnant.
The shown pair (PairB) merges within $\sim 80\,{\rm Myrs}$ after its properties are closest to those of \obj .
PairA is at a later stage of the quasar pairing than pairB, and merges in the simulation within $\sim 10\,{\rm Myrs}$. Both merging timescales are broadly consistent with the estimate of the \obj\ dynamical friction time (Equations \ref{eq:t_galfric} and \ref{eq:t_bhfric}).

\subsection{Implications for Gravitational Waves}

After dynamical friction, \obj\ is expected to undergo hardening via loss-cone scattering of nearby stars, then GW-dominated evolution in the pulsar-timing array (PTA) band. Given that the virial SMBH mass estimates for the two quasars are comparable, dynamical friction should proceed without the binary stalling at $\sim$10--100 pc\cite{dosopoulou_dynamical_2017}. Regardless of the binary mass ratio, tidal interactions between the nuclear star clusters may accelerate the binary’s orbital decay\cite{ogiya_accelerated_2020}. Depending on how gas-rich the resulting nucleus is, the sub-pc SMBH binary’s orbital evolution may be affected by circumbinary-disk accretion at sub-pc separations. Depending on how long this phase lasts and how massive the gas supply is, the binary may also accrete an appreciable amount of mass, which in turn decreases the GW coalescence timescale\cite{siwek_effect_2020}. 
\par
Various PTAs have observed a common-spectrum stochastic process in their data that may be consistent with a stochastic GW background\cite{arzoumanian_nanograv_2020,chen_common-red-signal_2021,goncharov_evidence_2021}. Hellings-Downs correlations\cite{Hellings1983}, however, have not yet been observed, which are needed in order to determine that this signal is of GW origin. If the common-process signal that current PTAs are observing is assumed to be due to a stochastic GW background and if the cosmic SMBH binary population at nanohertz GW frequencies is assumed to be the sole source of this background, then the amplitude of this signal may be used to constrain the cosmic merger rate of SMBH binaries.
This can then be compared to the observed occurrence rate of dual AGN from current and future wide-area deep surveys. 
The growing number of observations of dual AGNs/quasars across cosmic time as well as PTAs improving their characterization of a potential stochastic GW background will continue to provide vital tests for astrophysical models of SMBH binary formation.






%
%
\newgeometry{margin=2cm}
\begin{landscape}
\begin{table}
\centering
\begin{tabular}{ccccccccccc}
  \hline
  \hline
& & F475W & F814W & F160W$_{{\rm PSF}}$ & $K_{p, {\rm PSF}}$ & F160W$_{{\rm host}}$ & $K_{p, {\rm host}}$ & log$M_{\ast}$ & log$M_{\rm BH}$ & log$M_{{\rm BH,\,vir}}$ \\
Nucleus & Redshift & (mag) & (mag) & (mag) & (mag) & (mag) & (mag) & ($M_{\odot}$) & ($M_{\odot}$) & ($M_{\odot}$)  \\
(1) & (2) & (3) & (4) & (5) & (6) & (7) & (8) & (9) & (10) & (11) \\
\hline
\obj\ NE\dotfill & 2.1665 $\pm$ 0.0010 & 21.08 & 20.56 & 20.61 & 20.85 & 21.35 & 20.75 &  11.46  & 9.07  & 9.12  \\
\obj\ SW\dotfill & 2.1679 $\pm$ 0.0003 & 19.48 & 19.19 & 19.34 & 19.62 & 21.05 & 20.51 &  11.50  & 9.12 & 9.20 \\
\hline
\end{tabular}
\caption*{\textbf{Supplementary Table 1:} {\bf Basic photometric and spectroscopic properties of \obj}. Col. 1: NE: northeast nucleus. SW: southwest nucleus; Col. 2: Spectroscopic redshift measured from the best-fit model of the joint optical-NIR spectral fitting (Supplementary Figure 6); Cols. 3 \& 4: HST optical dual-band AB magnitude from \cite{ChenHwang2022}; Cols. 5 \& 6: HST and Keck NIR PSF AB magnitude; Cols. 7 \& 8: HST and Keck NIR model AB magnitude for the extended host galaxies. Col. 9: Total host galaxy stellar mass estimate derived from $K_{p,\,{\rm host}}$ and the F160W$_{{\rm host}}$ - $K_{p,\,{\rm host}}$ color (Supplementary Figure 10). Col. 10: Black hole mass estimate inferred from host galaxy total stellar mass assuming the $M_{\rm BH}$--$M_{\ast}$ relation observed in local \hst -imaged broad-line AGNs with virial BH mass estimates\cite{Bennert21}. Col. 11: Virial black hole mass estimate inferred from broad \halpha\ assuming the calibration of \cite{Shen2012}. }
\label{tab:basic_info}
\end{table}
\end{landscape}
\restoregeometry

\begin{table}
\centering
\begin{tabular}{lcc}
\hline\hline
~~~~~~Emission-line Measurements~~~~~~ & \obj\ NE &  \obj\ SW \\
\hline
\CIII\ Flux (10$^{-17}$ erg s$^{-1}$ cm$^{-2}$)         \dotfill(1) & 246$\pm$13 & 736$\pm$16   \\
\MgII\ Flux (10$^{-17}$ erg s$^{-1}$ cm$^{-2}$)         \dotfill(2) & 110$\pm$4 & 466$\pm$8   \\
\hbeta\ Flux (10$^{-17}$ erg s$^{-1}$ cm$^{-2}$)         \dotfill(3) & 51$\pm$14 & 233$\pm$27   \\
\OIIIb\ Flux (10$^{-17}$ erg s$^{-1}$ cm$^{-2}$)         \dotfill(4) & 8$\pm$4 & 22$\pm$9   \\
Broad \halpha\ Flux (10$^{-17}$ erg s$^{-1}$ cm$^{-2}$) \dotfill(5) & 151$\pm$10 & 309$\pm$21   \\
Narrow \halpha\ Flux (10$^{-17}$ erg s$^{-1}$ cm$^{-2}$) \dotfill(6) & 8$\pm$3 & 21$\pm$10   \\
\CIII\ EW ({\rm \AA })         \dotfill(7) & 29$\pm$2 & 22$\pm$1  \\
\MgII\ EW ({\rm \AA })         \dotfill(8) & 28$\pm$1 & 32$\pm$1   \\
\hbeta\ EW ({\rm \AA })         \dotfill(9) & 45$\pm$14 & 81$\pm$13    \\
\OIIIb\ EW ({\rm \AA })         \dotfill(10) & 7$\pm$4 & 8$\pm$3   \\
Broad \halpha\ EW ({\rm \AA })         \dotfill(11) & 181$\pm$16 & 150$\pm$14   \\
Narrow \halpha\ EW ({\rm \AA })         \dotfill(12) & 9$\pm$4 & 10$\pm$5   \\
\CIII\ FWHM (km s$^{-1}$)    \dotfill(13) & 8480$\pm$1200 & 7900$\pm$290   \\
\MgII\ FWHM (km s$^{-1}$)    \dotfill(14) &  3980$\pm$160 & 4450$\pm$100   \\
\hbeta\ FWHM (km s$^{-1}$)    \dotfill(15) & 7600$\pm$1800 & 7400$\pm$1000   \\
\OIIIb\ FWHM (km s$^{-1}$)    \dotfill(16) & 910$\pm$410 & 680$\pm$200  \\
Broad \halpha\ FWHM (km s$^{-1}$)    \dotfill(17) & 5550$\pm$400 & 4680$\pm$430   \\
\hline
\end{tabular}
\caption*{\textbf{Supplementary Table 2:} {\bf Emission-line measurements.} The measurements are from jointly analyzing the Gemini/GMOS and Gemini/GNIRS spectra.
Lines 1--6: emission-line flux intensity. For \MgII, only the broad component is listed here. 
Lines 7--12: rest-frame equivalent width.  
Lines 13--17: Full width at half maximum. All errors quoted are 1-$\sigma$ uncertainties estimated from Monte Carlo simulations.
}
\label{tab:spectral_measurement}
\end{table}

\newgeometry{margin=2cm}
\begin{landscape}
\begin{table}
\begin{center}
\begin{tabular}{ccccccccc}
\hline
\hline
& & flux$_{0.5-2\,{\rm keV}}$ & flux$_{2-8\,{\rm keV}}$ & & & N$_{{\rm H}}$ & log $L_{0.5-2\,{\rm keV}}$ & log $L_{2-8\,{\rm keV}}$  \\
Nucleus & Counts & \multicolumn{2}{c}{(10$^{-6}$ photons cm$^{-2}$ s$^{-1}$)}  & HR & $\Gamma_{X}$ & (10$^{22}$ cm$^{-2}$) & (erg s$^{-1}$) & (erg s$^{-1}$)   \\
(1) & (2) & (3) & (4) & (5) & (6) & (7) & (8) & (9)  \\
\hline
\obj\ NE & 191.5 & 12.26$^{+3.71}_{-1.06}$ & 10.52$^{+0.24}_{-0.48}$ & 0.37$^{+0.01}_{-0.02}$ & 1.8 & 4.7$^{+1.3}_{-2.6}$ & 44.92$^{+0.02}_{-0.03}$ & 45.29$^{+0.05}_{-0.03}$   \\
\obj\ SW & 6 & 0.56$^{+0.51}_{-0.41}$ & 0.23$^{+0.21}_{-0.17}$ & $-$0.34$^{+0.48}_{-0.60}$ & 1.8 & $<$10$^{-2}$ & 43.17$^{+0.32}_{-0.60}$ & 43.62$^{+0.32}_{-0.58}$ \\
\hline
\end{tabular}
\caption*{\textbf{Supplementary Table 3:} {\bf X-ray measurements from Chandra ACIS-S observations.} Col. 2: Total net counts in 0.5--8 keV band. Cols. 3 \& 4: Observed photon flux. 
Col. 5: Hardness ratio HR$\equiv(H-S)/(H+S)$ where $H$ and $S$ are the number of counts in the hard and soft X-ray bands, respectively. Col. 6: Photon index value (fixed), assuming a power-law model where $n(E)\propto E^{-\Gamma_X}$. Col. 7: Best-fit intrinsic column density assuming a power-law model. Cols. 8 \& 9: Unabsorbed luminosity. All values and errors represent the median value and 99.7\% confidence interval when analyzing 100 spectral realizations of each X-ray point source.   
}\label{tab:xray}
\end{center}
\end{table}
\end{landscape}
\restoregeometry

\newgeometry{margin=1cm}
\begin{table}
\centering
{
\begin{tabular}{lcc}
  \hline
  \hline
  & \obj\ NE & \obj\ SW \\
  \hline
  \hline
 C-band (6.0 GHz) \\
 \hline
 C (hh:mm:ss.ss+dd:mm:ss.s) \dotfill(1) & 07:49:22.99+22.55.12.0 & 07:49:22.96+22.55.11.7\\
 S$_{{\rm Peak}}$ (mJy/beam) \dotfill(2) & 22.91$\pm$0.05 & 60.43$\pm$0.05\\
 T$_{{\rm Peak}}$ (K) \dotfill(3) & 8420$\pm$20 & 22200$\pm$20\\
 S$_{{\rm Int}}$ (mJy) \dotfill(4) & 22.94$\pm$0.05 & 60.51$\pm$0.05\\
 log $\nu$L$_{{\rm \nu,6.0\,GHz}}$ (erg s$^{-1}$)\dotfill(5) & 43.382$\pm$0.002 & 43.763$\pm$0.001\\
 ${\rm \theta_{maj}}$$\times$${\rm \theta_{min}}$ ($''$$\times$$''$) \dotfill(6) & \multicolumn{2}{c}{0\farcs44$\times$0\farcs21} \\
 PA ($^{\circ}$) \dotfill(7)& \multicolumn{2}{c}{70.9}\\
  \hline
  Ku-band (15.0 GHz) \\ 
  \hline
 C (hh:mm:ss.ss+dd:mm:ss.s) \dotfill(8)& 07:49:22.99+22.55.12.0 & 07:49:22.96+22.55.11.7\\
 S$_{{\rm Peak}}$ (mJy/beam) \dotfill(9) & 15.64$\pm$0.08 & 44.66$\pm$0.08\\
 T$_{{\rm Peak}}$ (K) \dotfill(10) & 7720$\pm$40 & 22050$\pm$40\\
 S$_{{\rm Int}}$ (mJy) \dotfill(11)& 15.95$\pm$0.14 & 45.38$\pm$0.14\\
 log $\nu$L$_{{\rm \nu,15.0\,GHz}}$ (erg s$^{-1}$) \dotfill(12) & 43.622$\pm$0.004 & 44.036$\pm$0.002 \\
 ${\rm \theta_{maj}}$$\times$${\rm \theta_{min}}$ ($''$$\times$$''$) \dotfill(13)& \multicolumn{2}{c}{0\farcs11$\times$0\farcs10 } \\
 PA ($^{\circ}$) \dotfill(14)& \multicolumn{2}{c}{$-$54.0}\\

\hline
\end{tabular}
\caption*{{\textbf{Supplementary Table 4:} {\bf Radio measurements.} The measurements are based on 2D Gaussian fits from VLA A-config dual-band imaging. Col. 1: J2000 coordinates for the best-fit centers in 6 GHz. Col. 2: 6.0 GHz peak flux density. Col. 3: 6.0 GHz peak brightness temperature. Col. 4: 6.0 GHz integrated flux density from decomposing the two marginally resolved nuclei. Col. 5: Rest-frame 6.0 GHz luminosity density. Col. 6: Semi-major and semi-minor axis of the clean beam. Col. 7: Position angle of the clean beam.
Col. 8-14: Similar to above, but for 15 GHz.  All errors quoted represent 1$\sigma$ statistical uncertainties. As a comparison, the peak brightness temperatures from the archival VLBA X-band (8.4 GHz) image are 1.6$\times$10$^8$ K for the SW nucleus and 5.2$\times$10$^7$ K for the NE nucleus. All error bars are $1\sigma$.  }
}\label{tab:radio}
}
\end{table}
\restoregeometry

\newgeometry{margin=0.1cm}
\begin{table}
\centering
\begin{tabular}{lcc}
\hline\hline
~~~~~~HST (F160W) GALFIT Parameters~~~~~~ & \obj\ NE &  \obj\ SW \\
\hline\hline
Model 1 (2 PSFs + 2 S\'ersics): $\chi^2_{\nu} = $ 1.331 \\
\hline
m$_{{\rm PSF}}$ (AB)         \dotfill(1) & 20.605$\pm$0.002  &   19.344$\pm$0.019  \\
    m$_{\rm S\acute{e}rsic}$ (AB) \dotfill(2) & 21.349$\pm$ 0.034  &    21.053$\pm$0.095  \\
$R_e$ (kpc)\dotfill(3) & 4.29$\pm$0.07 & 1.44$\pm$0.28 \\
n\dotfill(4) & 0.91$\pm$0.03 & 5.20$\pm$1.32\\
\hline
Model 2 (2 PSFs + 1 S\'ersic): $\chi^2_{\nu} = $1.354 \\
\hline
m$_{{\rm PSF}}$ (AB)         \dotfill(5) & 20.613$\pm$0.003  &  19.260$\pm$0.001   \\
C$_{{\rm PSF}}$ (hh:mm:ss.ss+dd:mm:ss.s)\dotfill(6) & 07:49:23.00+22:55:12.0 & 07:49:22.97+22:55:11.8 \\
m$_{\rm S\acute{e}rsic}$ (AB) \dotfill(7) & \multicolumn{2}{c}{ 20.734$\pm$0.008 }     \\
C$_{\rm S\acute{e}rsic}$ (hh:mm:ss.ss+dd:mm:ss.s)\dotfill(8) & \multicolumn{2}{c}{ 07:49:22.98+22:55:11.9 }\\
$R_e$ (kpc)\dotfill(9) & \multicolumn{2}{c}{ 3.47$\pm$0.04 } \\
n\dotfill(10) & \multicolumn{2}{c}{ 1.38$\pm$0.04 } \\
\hline
Model 3 (2 PSFs + 4 S\'ersics): $\chi^2_{\nu} = $1.327 \\
\hline
m$_{{\rm PSF}}$ (AB)         \dotfill(11) & 20.613$\pm$0.003  &  19.373$\pm$0.014  \\
m$_{\rm S\acute{e}rsic,\,1}$ (AB) \dotfill(12) &  21.476$\pm$0.038 &  21.257$\pm$0.057  \\
m$_{\rm S\acute{e}rsic,\,2}$ (AB) \dotfill(13) &  23.794$\pm$0.876 &  22.186$\pm$0.157  \\
$R_{e,\,1}$ (kpc)\dotfill(14) & 3.89$\pm$0.09 & 1.94$\pm$0.29 \\
$R_{e,\,2}$ (kpc)\dotfill(15) & 9.94$\pm$13.91 & 0.81$\pm$0.14 \\
n$_1$\dotfill(16) & 1 & 4\\
B/T  \dotfill(17)  &  0.11$\pm$0.08 &  0.70$\pm$0.05 \\
\hline
\end{tabular}
\caption*{\textbf{Supplementary Table 5:} {\bf HST image decomposition.} GALFIT structural decomposition results for the \hst /WFC3 F160W image. Lines 1 \& 2: Best-fit component magnitude. 
Lines 3 \& 4: Best-fit S\'ersic component effective radius and index. 
Lines 5--10: Similar to above, but for a single S\'ersic for the foreground lens. Lines 6 \& 8: J2000 coordinates for the best-fit centers. Lines 11--17: Similar to above, but for two S\'ersic models for each host to attempt bulge-disk decomposition. The S\'ersic index is fixed at $n=1$ ($n=4$) for the disk (bulge) in each host to help break model degeneracy.
Line 17: Bulge-to-total luminosity ratio. All error bars are $1\sigma$. 
}
\label{tab:galfit_hst}
\end{table}
\restoregeometry

\newgeometry{margin=1cm}
\begin{table}
\centering
\begin{tabular}{lcc}
\hline\hline
~~~~~~Keck ($K_p$) GALFIT Parameters~~~~~~ & \obj\ NE &  \obj\ SW \\
\hline\hline
Model 1 (2 PSFs + 2 S\'ersics): $\chi^2_{\nu} = $1.416 \\
\hline
m$_{{\rm PSF}}$ (AB)         \dotfill(1) & 20.847$\pm$0.006 &  19.620$\pm$0.002 \\
m$_{\rm S\acute{e}rsic}$ (AB) \dotfill(2) & 20.745$\pm$0.017 & 20.514$\pm$0.008    \\
$R_e$ (kpc)\dotfill(3) &  1.76$\pm$0.04 &  1.07$\pm$0.01 \\
n\dotfill(4) & 0.56$\pm$0.05 &  0.16$\pm$0.02\\
\hline
Model 2 (2 PSFs + 1 S\'ersic): $\chi^2_{\nu} = $1.521\\
\hline
m$_{{\rm PSF}}$ (AB)         \dotfill(5) &  20.706$\pm$0.004 &  19.573$\pm$0.002   \\
C$_{{\rm PSF}}$ (hh:mm:ss.ss+dd:mm:ss.s)\dotfill(6) &  07:49:23.00+22:55:12.0 & 07:49:22.97+22:55:11.8\\
m$_{\rm S\acute{e}rsic}$ (AB) \dotfill(7) & \multicolumn{2}{c}{19.925$\pm$0.011}     \\
C$_{\rm S\acute{e}rsic}$ (hh:mm:ss.ss+dd:mm:ss.s)\dotfill(8) & \multicolumn{2}{c}{ 07:49:22.97+22:55:11.7 }\\
$R_e$ (kpc)\dotfill(9) & \multicolumn{2}{c}{ 1.79$\pm$0.04} \\
n\dotfill(10) & \multicolumn{2}{c}{1.405$\pm$0.04}\\
\hline
\end{tabular}
\caption*{\textbf{Supplementary Table 6:} {\bf Keck image decomposition.} GALFIT structural decomposition results for the Keck/NIRC2 AO $K_p$ image. Lines 1 \& 2: Best-fit component magnitude. 
Lines 3 \& 4: Best-fit S\'ersic component effective radius and index. 
Lines 5--10: Similar to above, but for a single S\'ersic model for the foreground lens. Lines 6 \& 8: J2000 coordinates for the best-fit centers. The absolute astrometry of the Keck image has been registered to that of the \hst\ image according to the PSF centers. All error bars are $1\sigma$. 
}
\label{tab:galfit_keck}
\end{table}
\restoregeometry

\begin{table}
\centering
\begin{tabular}{lcccc}
\hline\hline
Identifier & $n_{\rm pix}$ &  mag/mag uncertainty & significance level & location \\
\hline
T1 & 47 & 25.47 $\pm$ 0.07 & 23 & SW to the pair   \\
T2 & 45 & 25.92 $\pm$ 0.07 & 15 & S to the pair   \\
T3 & 35 & 26.22 $\pm$ 0.06 & 13 & N to the pair   \\
T4 & 30 & 26.50 $\pm$ 0.06 & 11 & E to the pair   \\
\hline
\end{tabular}
\caption*{
\textbf{Supplementary Table 7: Tidal features.} Identified tidal features from the HST residual map and their flux (AB magnitude) measurements. All error bars are $1\sigma$. 
}
\label{tab:tidal}
\end{table}


\begin{table}
\centering
{
\begin{tabular}{lccc}
\hline\hline
Strong Lensing Mass Model Parameters & SIE (\hst ) &  SIE (Keck) & SIE+shear (Keck) \\
\hline
$\chi^2_{{\rm min}}$\dotfill(1) & $\sim$0 & 39.9 & 3.2  \\
$\theta_{{\rm Ein}}$ ($''$)\dotfill(2) & 0.24$\pm$0.01 & 0.23$\pm$0.01 & 0.13$\pm$0.03  \\
$\Delta_{{\rm RA}}$ ($''$)\dotfill(3) & 0.25$\pm$0.03 & 0.22$\pm$0.02 & 0.07$\pm$0.02   \\
$\Delta_{{\rm Dec}}$ ($''$)\dotfill(4) & 0.12$\pm$0.03 & 0.10$\pm$0.02 & 0.04$\pm$0.01   \\
$e$\dotfill(5) & 0.17$\pm$0.07 & 0.06$\pm$0.03 & 0.23$\pm$0.14   \\
PA$_e$ ($^{\circ}$)\dotfill(6) & 37$\pm$11 & 41$\pm$8 & 56$\pm$17   \\
shear\dotfill(7) & N/A & N/A & 0.48$\pm$0.12  \\
PA$_{\rm shear}$ ($^{\circ}$)\dotfill(8) & N/A & N/A & $-$35$\pm$3   \\
\hline
\end{tabular}
}
\caption*{
{
\textbf{Supplementary Table 8: Lens modeling.} Strong lensing mass modeling parameters (mean values and 1-$\sigma$ uncertainties except for the minimum $\chi^2$) obtained using glafic from the MCMC analysis for the \hst\ (assuming the SIE model) and Keck (assuming the SIE or SIE+shear model) images.
Line 1: Minimum $\chi^2$ given by the best-fit model.
Line 2: Einstein radius.  
Lines 3 \& 4: Lens galaxy position with respect to the SW quasar image.  
Line 5: Ellipticity of the ellipsoid. $e=0$ for a spherical case. Line 6: Position angle of the ellipsoid measured East of North. Line 7: External shear. Line 8: Position angle of the external shear measured East of North. All errors quoted are 1$\sigma$ uncertainties estimated from the MCMC analysis.
}
}
\label{tab:lensing}
\end{table}

\clearpage


\begin{figure}
    \centering
    \includegraphics[width=0.8\textwidth]{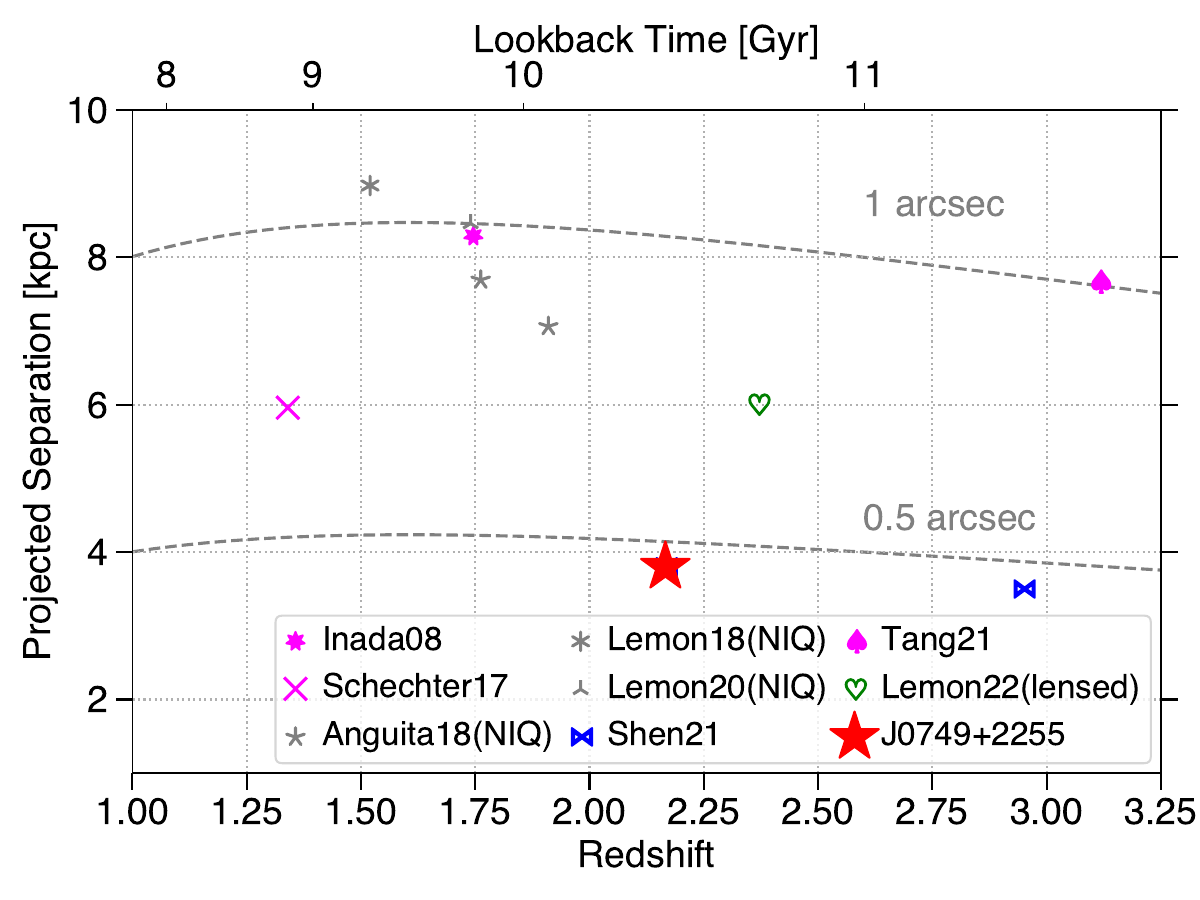}
    \caption*{
    \small
    \textbf{Supplementary Figure 1: Pair separation versus redshift. }
    \obj\ in comparison to the previously known candidate dual quasars in the literature at $z>1$ and with projected separations $<$10 kpc. Nearly identical quasars (NIQs)\cite{Anguita2018,Lemon2018,Lemon2020}, shown with grey symbols, are marked with "(NIQ)" in the legend. They are generally thought to be lensed quasars lacking a definitive detection of foreground lenses due to observational limitations. Three other dual quasar candidates have more noticeable spectral differences\cite{Inada2008,Schechter2017,Tang2021} (shown in magenta), but they do not show unambiguous host merger features, and are debatable as project pairs at much larger separations or lenses. The dual quasar candidate of \cite{Shen2021} (shown in blue) shows differences in the broad emission line strengths\cite{Mannucci2022} but is still controversial as a quasar lens due to the lack of high-resolution deep NIR imaging. The quadruply lensed quasar\cite{Lemon2022} (shown in green) has been suggested to contain a lensed kpc-scale dual quasar, but the interpretation is subject to lense modeling. {There are two more $z>1$ candidate dual AGNs not shown given their much smaller separations: a strongly lensed binary AGN candidate at $z=2.51$ in PKS 1830-211 with sub-parsec separation\cite{Nair2005} and a strongly lensed dual AGN candidate\cite{Spingola2019,Schwartz2021} at $z=3.273$ in MG B2016+112 with a separation of hundreds of parsec.}
    \normalsize
    }
    \label{fig:z-sep}
\end{figure}

\begin{figure}
  \centering
    \includegraphics[width=1.0\textwidth]{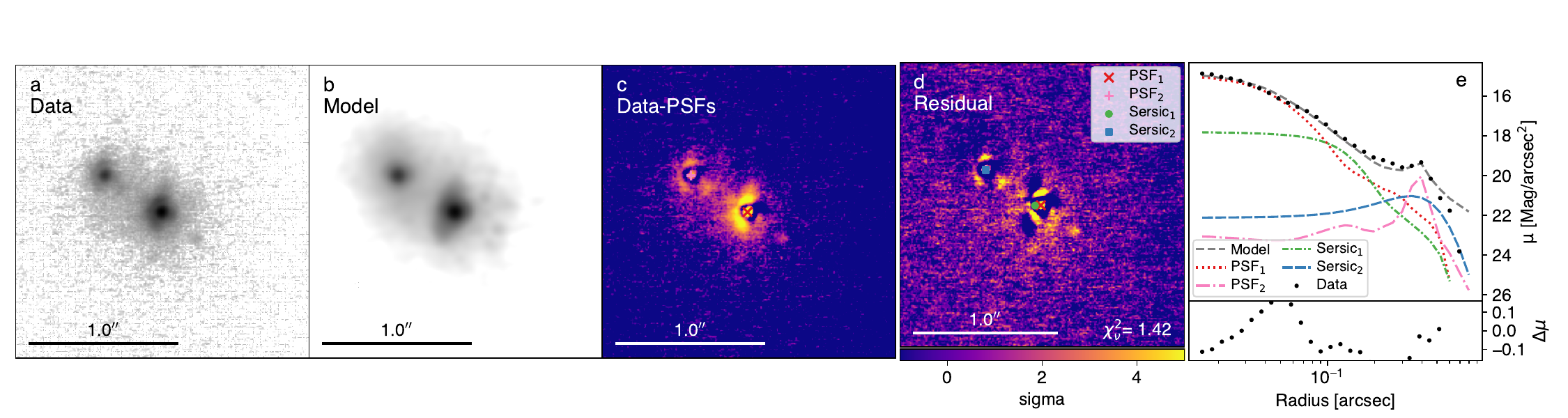}
    \caption*{
    \textbf{Supplementary Figure 2: Keck imaging.}
    Similar to Figure \ref{fig:hst_galfit}, but for Keck AO-assisted $K_p$ imaging of \obj\ and 2D structural decomposition results from GALFIT analysis. The best-fit GALFIT model contains two PSFs for the two unresolved quasars and two S\'{e}rsic models for the two host galaxies with the $\chi^2_{\nu}$ value labeled on the residual image. After we subtract the two PSFs, the extended host galaxies are detected (i.e., Data-PSFs). The tidal features detected in the \hst\ image (in the GALFIT model residuals) are too faint to detect in the shallower Keck image. The lensing hypotheis is disfavored based on both the $\chi^2_{\nu}$ values and lensing mass model tests. 
    }
    \label{fig:keck_galfit}
\end{figure}

\begin{figure}
  \centering
    \includegraphics[width=1.0\textwidth]{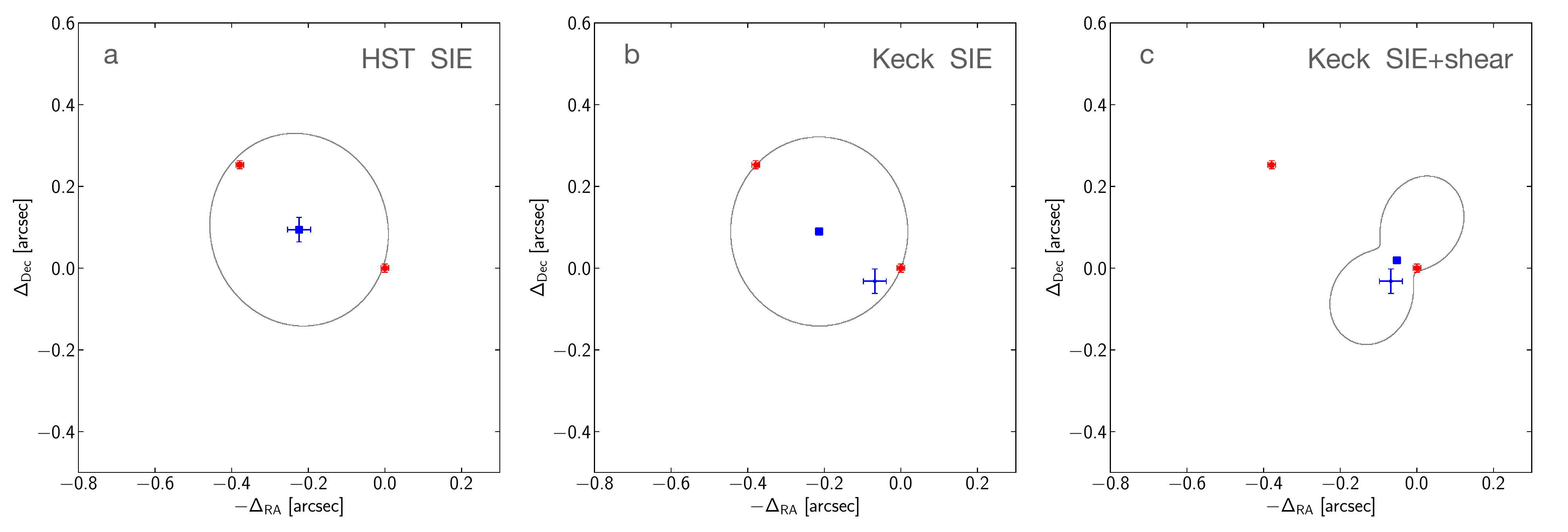}
    \caption*{
    \textbf{Supplementary Figure 3: Strong Lensing Mass Models.}
   Strong lensing tests from glafic. Panel a and panel b show the strong lensing mass model results using a singular isothermal ellipsoid (SIE) lens based on the HST and Keck images, respectively. Panel c shows a  more complicated SIE plus shear model from the Keck image. The solid lines are the critical curves.   The filled circles with error bars indicate the positions of quasars (red) and the putative lensing galaxy (blue) from HST and Keck images. The filled blue squares are the positions of the lensing galaxy predicted by the best-fitting mass models. 
    }
    \label{fig:glafic}
\end{figure}

\begin{figure}
    \centering
    \includegraphics[width=0.3\textwidth]{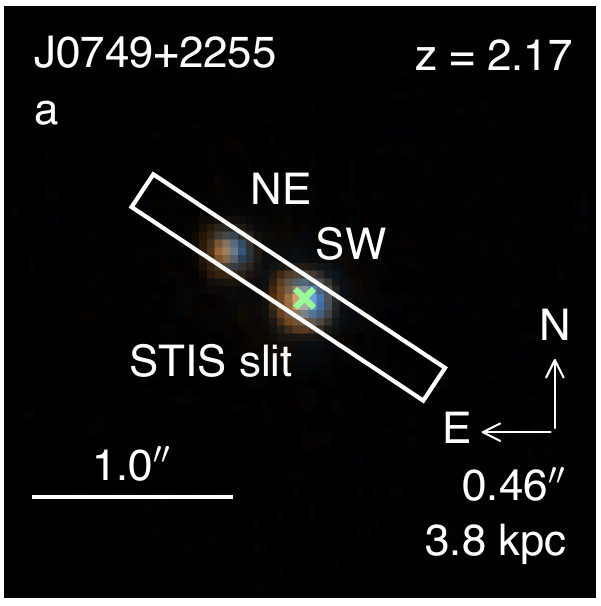}
    \includegraphics[width=0.60\textwidth]{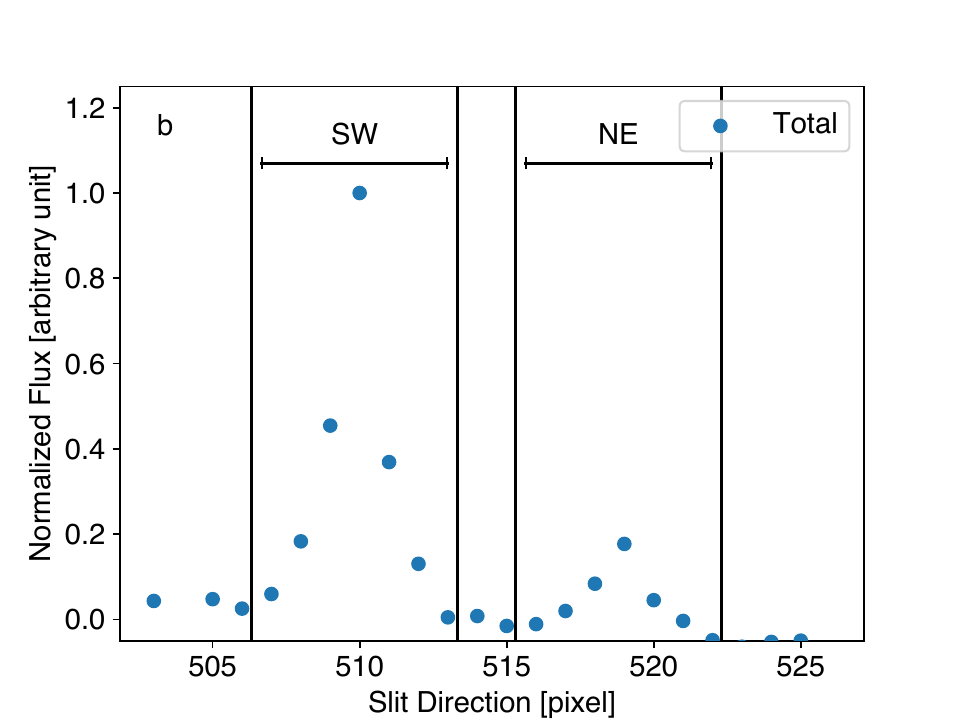}
    \includegraphics[width=0.85\textwidth]{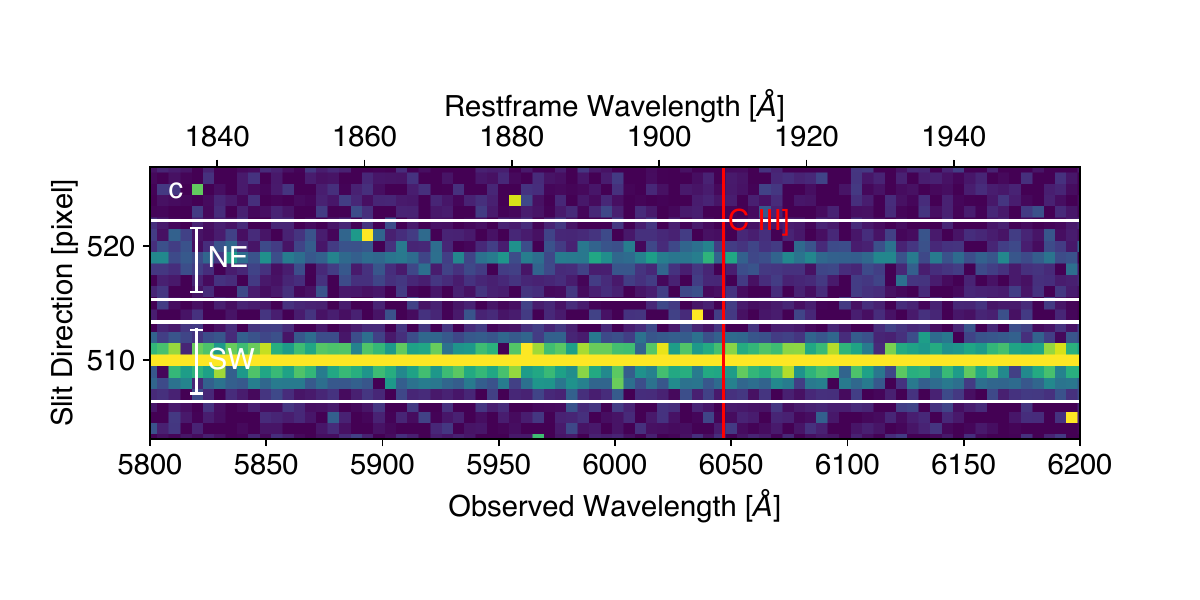}
    \caption*{
    \textbf{Supplementary Figure 4: HST/STIS spectrum.} 
    2D HST/STIS spectrum for \obj\ and illustration of the slit position and 1D aperture extraction for each nucleus. Panel a: Illustration of the STIS slit position and orientation. The background shows the \hst\ optical two-band (F475W and F814W) color image\cite{Shen2021}. The width of the white rectangle denotes the slit width. Panel b: Projected 1D spatial profile of the wavelength-averaged STIS spectrum. The blue filled circles denote the total spectrum. The black vertical lines mark the extraction apertures for the two nuclei. Panel c: 2D STIS spectrum around the C\,{\small III]} line. The white horizontal lines mark the extraction apertures for the two nuclei.
    }
    \label{fig:STIS_2d}
\end{figure}

\begin{figure}
    \centering
    \includegraphics[width=0.3\textwidth]{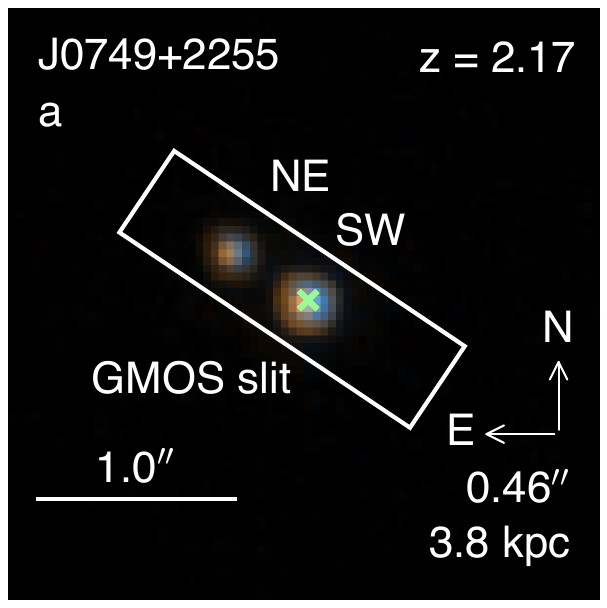}
    \includegraphics[width=0.60\textwidth]{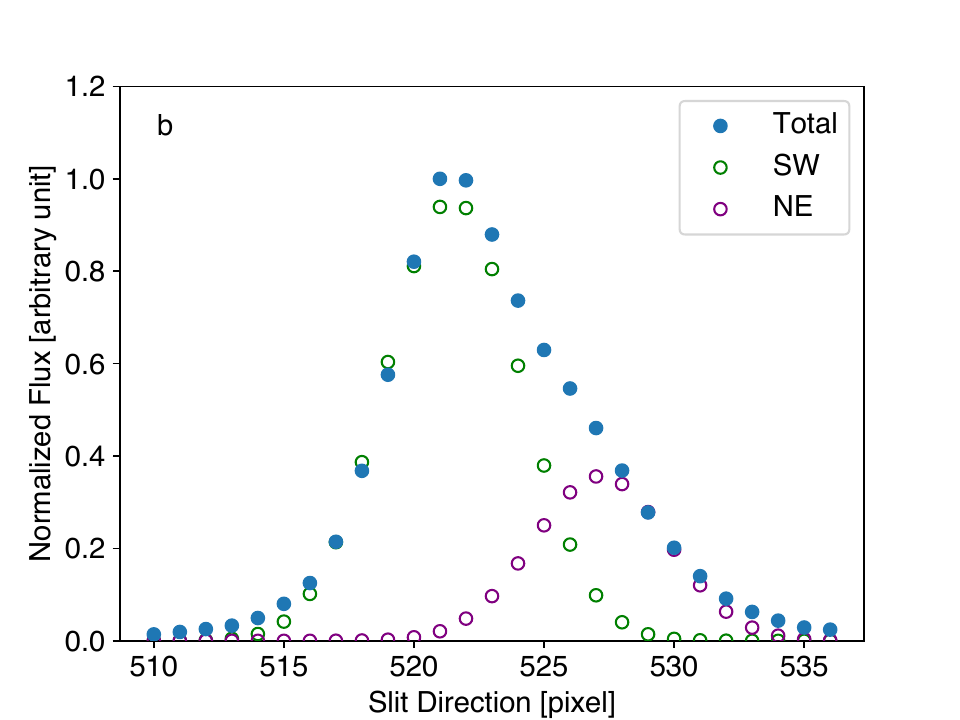}
    \includegraphics[width=0.85\textwidth]{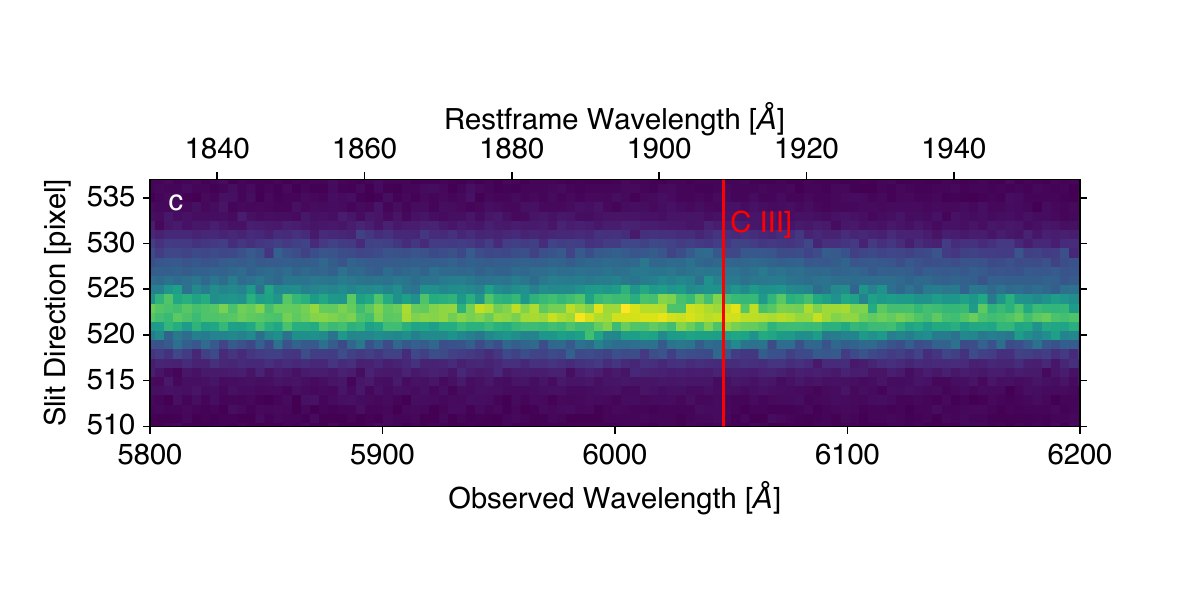}
    \caption*{
    \textbf{Supplementary Figure 5: Gemini/GMOS spectrum.} 
    2D Gemini/GMOS spectrum for \obj\ and illustration of the slit position and source decomposition for each nucleus. Panel a: Illustration of the GMOS slit position and orientation. The background shows the \hst\ optical two-band (F475W and F814W) color image\cite{Shen2021}. The width of the white rectangle denotes the slit width. Panel b: Projected 1D spatial profile of the wavelength-averaged GMOS spectrum. The blue filled circles denote the total spectrum. The green (purple) open circles represent the decomposed spatial profile of the SW (NE) nucleus. Panel c: 2D GMOS spectrum around the C\,{\small III]} line.}
    \label{fig:GMOS_2d}
\end{figure}

\newgeometry{margin=1cm}
\begin{figure}
    \centering
    \includegraphics[width=0.50\textwidth]{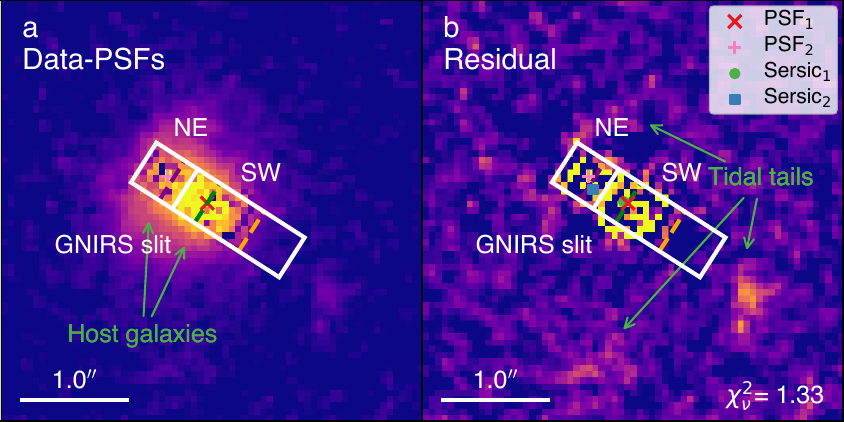}
    \includegraphics[width=0.55\textwidth]{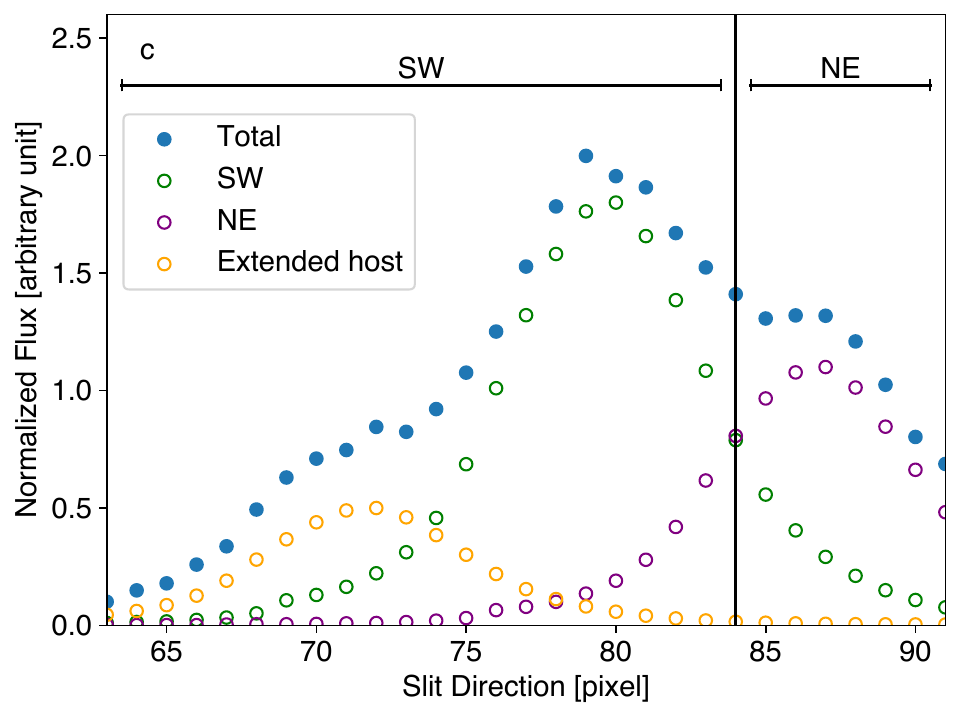}
    \includegraphics[width=0.35\textwidth]{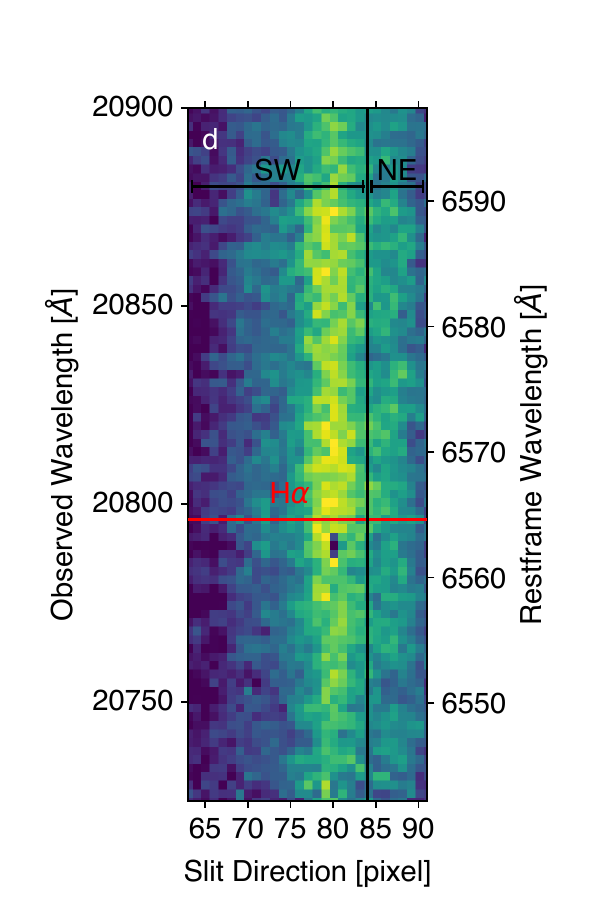}
    \caption*{
    \textbf{Supplementary Figure 6: Gemini/GNIRS spectrum. } 
    2D Gemini/GNIRS spectrum for \obj\ and illustration of the slit position and source decomposition for each nucleus. Panels a \& b: The GNIRS slit position and orientation. The background shows the \hst\ F160W PSF-subtracted (Panel a) and residual (Panel b) images. The white rectangles marked with ``NE'' and ``SW'' show the extraction apertures of both nuclei. The three dashed lines mark the peak positions of three components seen in the projected 1D profile. Panel c: Projected 1D spatial profile of the wavelength-average GNIRS spectrum. The blue filled circles show the total spectrum. The open circles represent the SW nucleus (shown in green), the NE nucleus (purple), and the extended host (orange). The top horizontal black lines mark the extraction apertures for both nuclei. Panel d: 2D GNIRS spectrum around the H$\alpha$ line. The horizontal black lines denote the extraction apertures for both nuclei.}
    \label{fig:GNIRS_2d}
\end{figure}
\restoregeometry

\newgeometry{margin=1cm}
\begin{figure}
  \centering
    \includegraphics[width=0.8\textwidth]{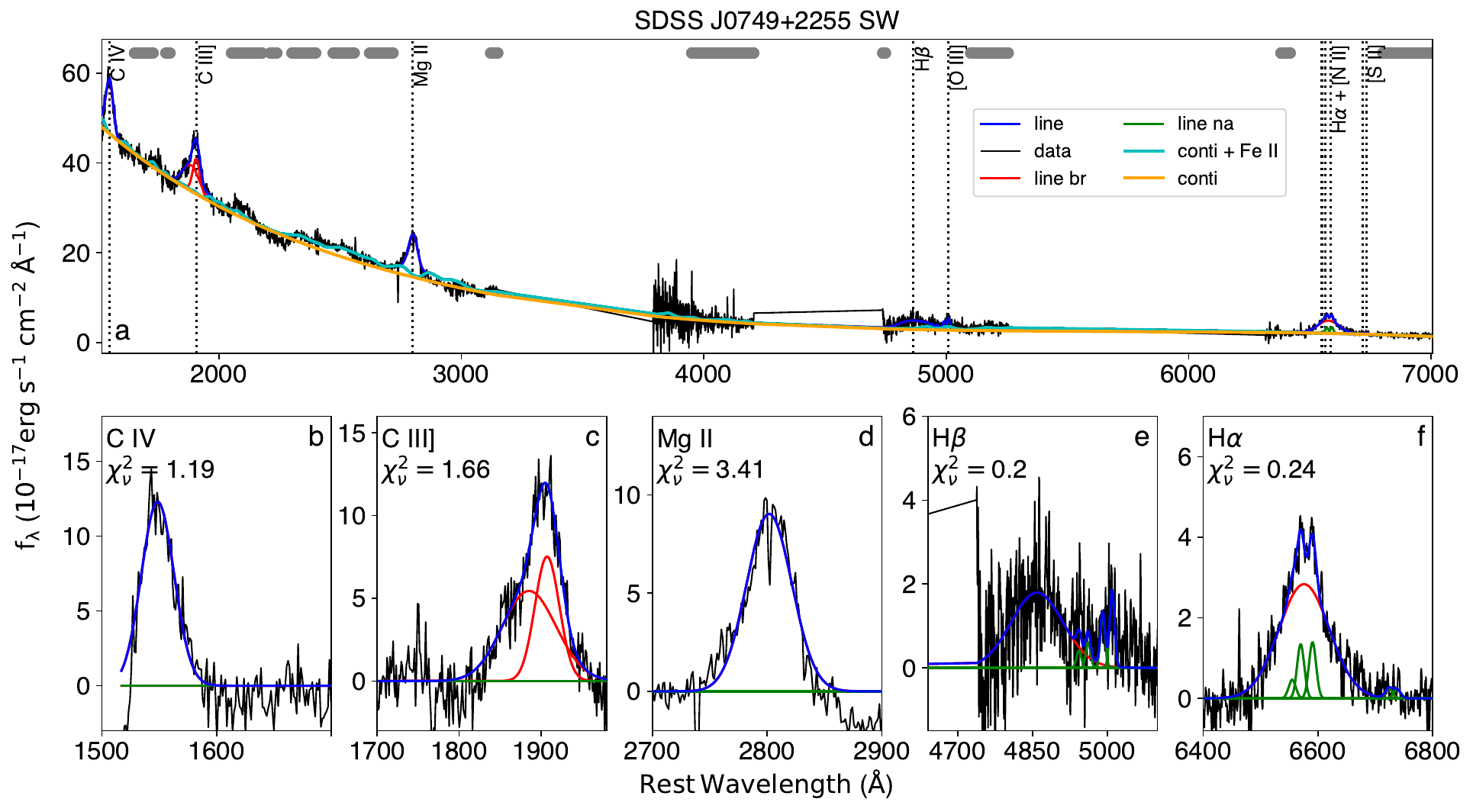}
    \includegraphics[width=0.8\textwidth]{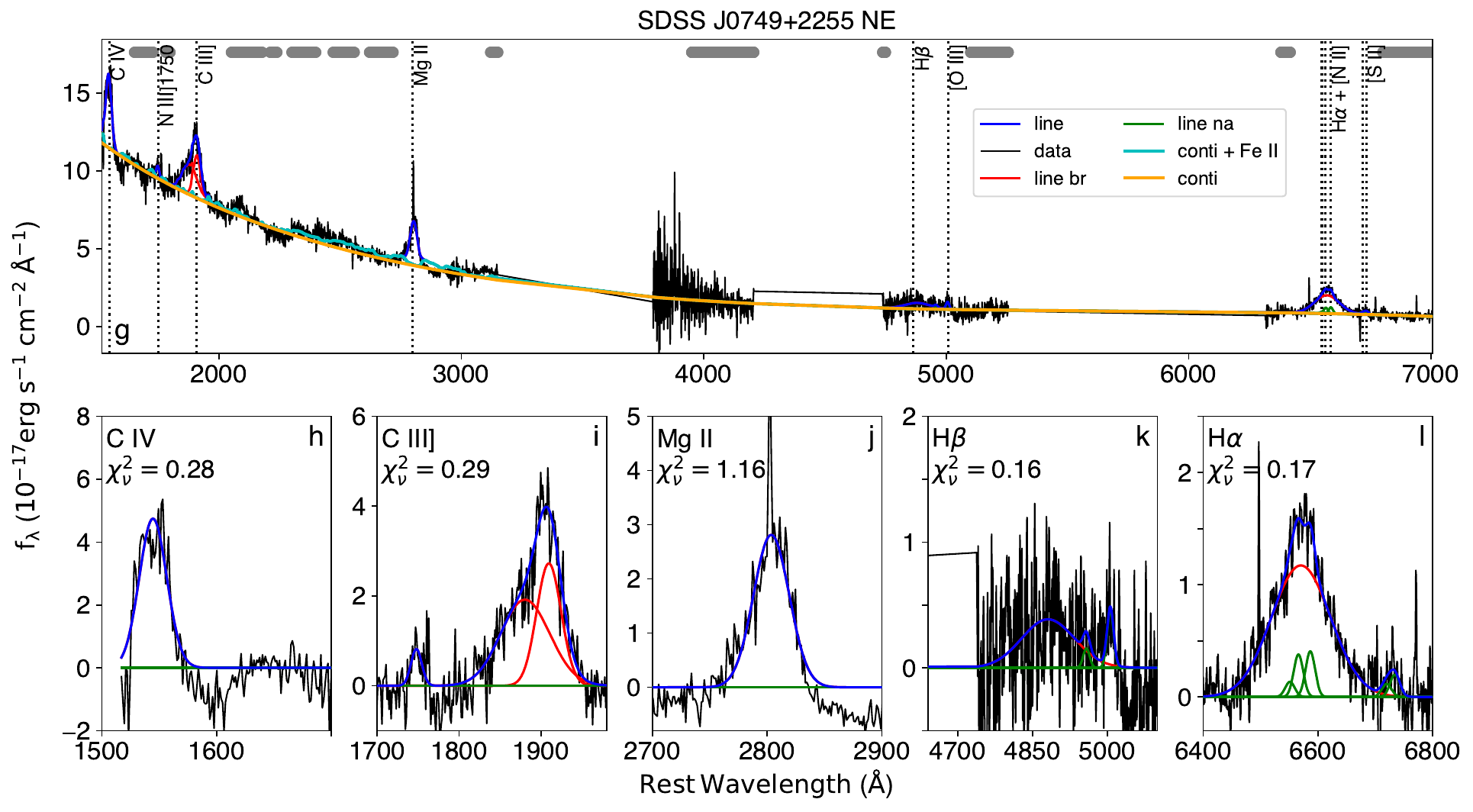}
    \caption*{
    \textbf{Supplementary Figure 7: Spectral fits.} 
    Global and local spectral fitting analysis for the joint optical-NIR spectrum of the two nuclei in \obj . Panels a--f (Panels g--l) show results for the SW (NE) nuclei. Panels a \& g: Global fit of the pseudo-continuum (in cyan), which consists of a power-law plus polynomial (in orange), and a fit with the Fe II template. The horizontal grey bars denote the wavelength windows adopted in the pseudo-continuum fit. Panels b--f, h--l: Local fit for individual emission lines (with the $\chi^2_{\nu}$ values marked) including C {\small IV} through H$\alpha$ using the pseudo-continuum-subtracted spectra. Red curves show models for the broad emission lines, whereas the green and blue curves denote models for the narrow and total emission lines, respectively. 
    }
    \label{fig:specfit}
\end{figure}
\restoregeometry

\begin{figure}
  \centering
    \includegraphics[width=\textwidth]{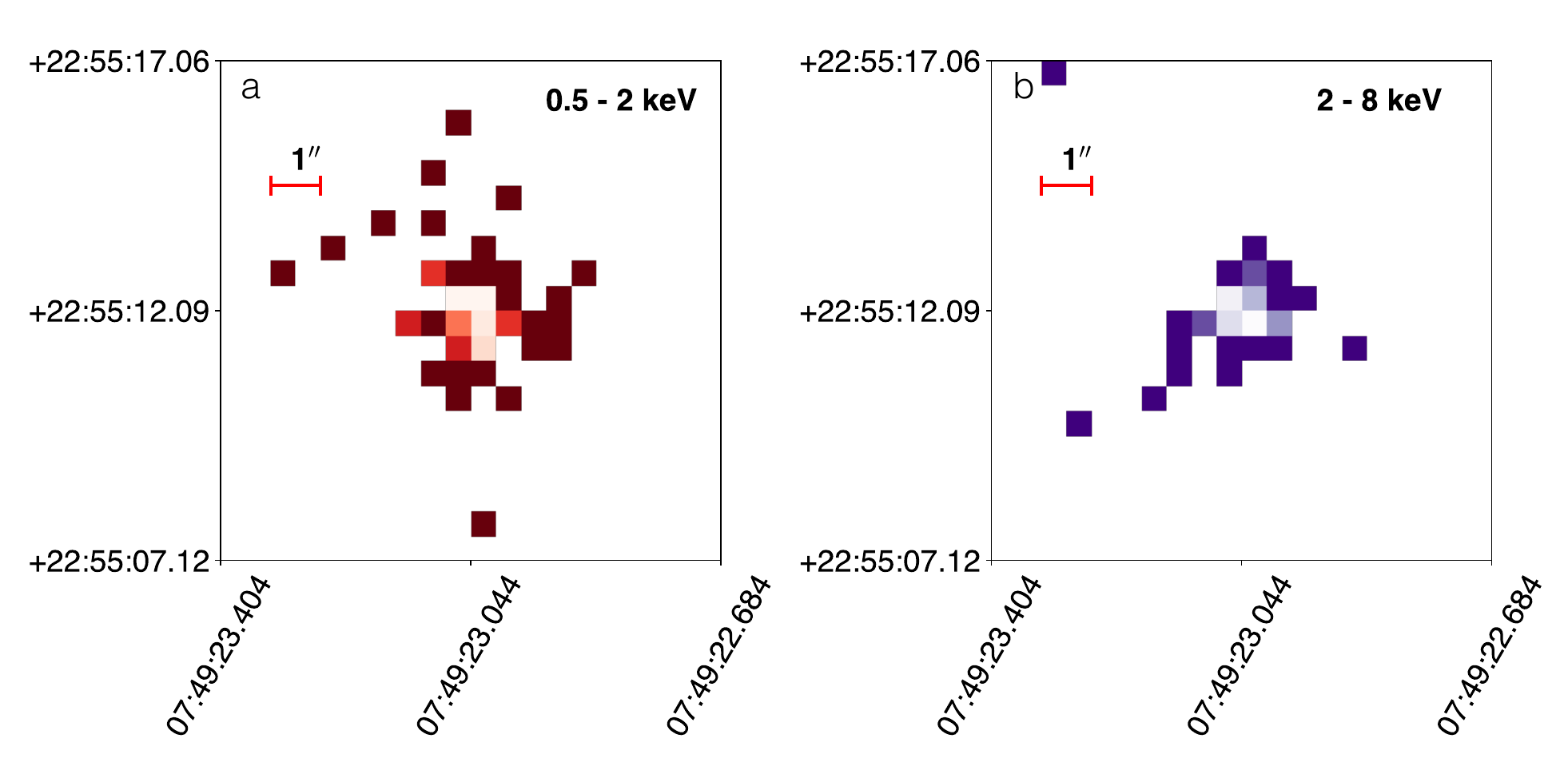}
    \caption*{
    \textbf{Supplementary Figure 8: X-ray analysis.} Binned \chandra~data, centered on the nominal X-ray coordinates of \obj. Panel a shows soft photons (0.5--2 keV) (total of 114 counts) and Panel b shows hard photons (2--8 keV) (total of 126 counts). In general, we find that the hardness ratio of the primary point source is consistent with an AGN ($HR =0.37^{+0.01}_{-0.02}$, while the secondary point source has a softer spectrum (albeit, with larger error bars; $HR=-0.34^{+0.48}_{-0.60}$)
    }
    \label{fig:xray_soft_hard}
\end{figure}

\begin{figure}
  \centering
    \includegraphics[width=\textwidth]{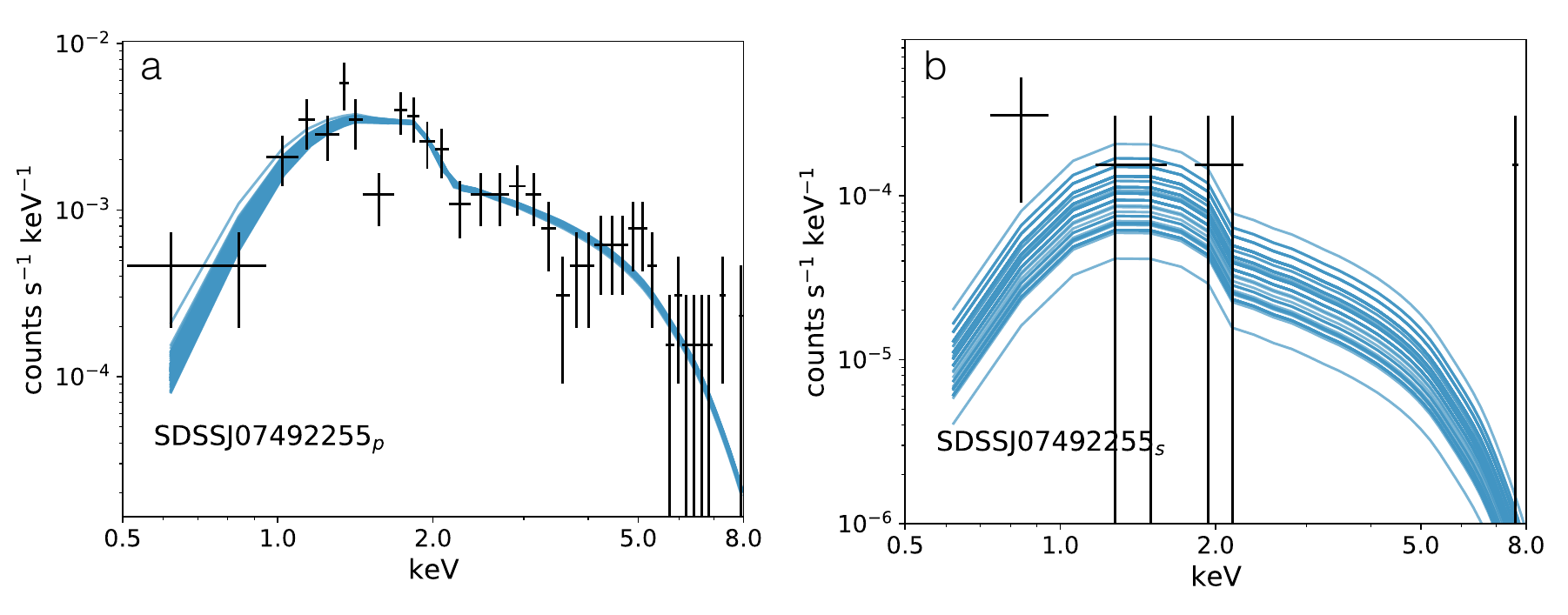}
    \caption*{
    \textbf{Supplementary Figure 9: X-ray spectral fits.} Spectral fits to the primary (Panel a) and secondary (Panel b) X-ray point source detected by {\tt BAYMAX}. We fit and model 100 spectral realizations for each X-ray point source (blue) using {\tt XSPEC} to quantify the average values of the spectral shape and X-ray luminosity. We show one of these spectral realizations in black, for each X-ray point source.  We find both sets of spectral realizations are best-fit with a simple absorbed power law ({\tt XSPEC} model components: {\tt phabs} $\times$ {\tt zphabs} $\times$ {\tt zpow}). Given the low number of counts for both components (average of $\sim191.5$ and $\sim6$ counts for the primary and secondary, respectively), we fix the power law index, $\Gamma$, to a value of 1.8 for all spectral fits. We list the median values and the confidence interval of the best-fit spectral parameter distributions in Supplementary Table 3. All error bars are 1$\sigma$.
    }
    \label{fig:xray_specfit}
\end{figure}

\begin{figure}
    \centering
    \includegraphics[width=0.8\textwidth]{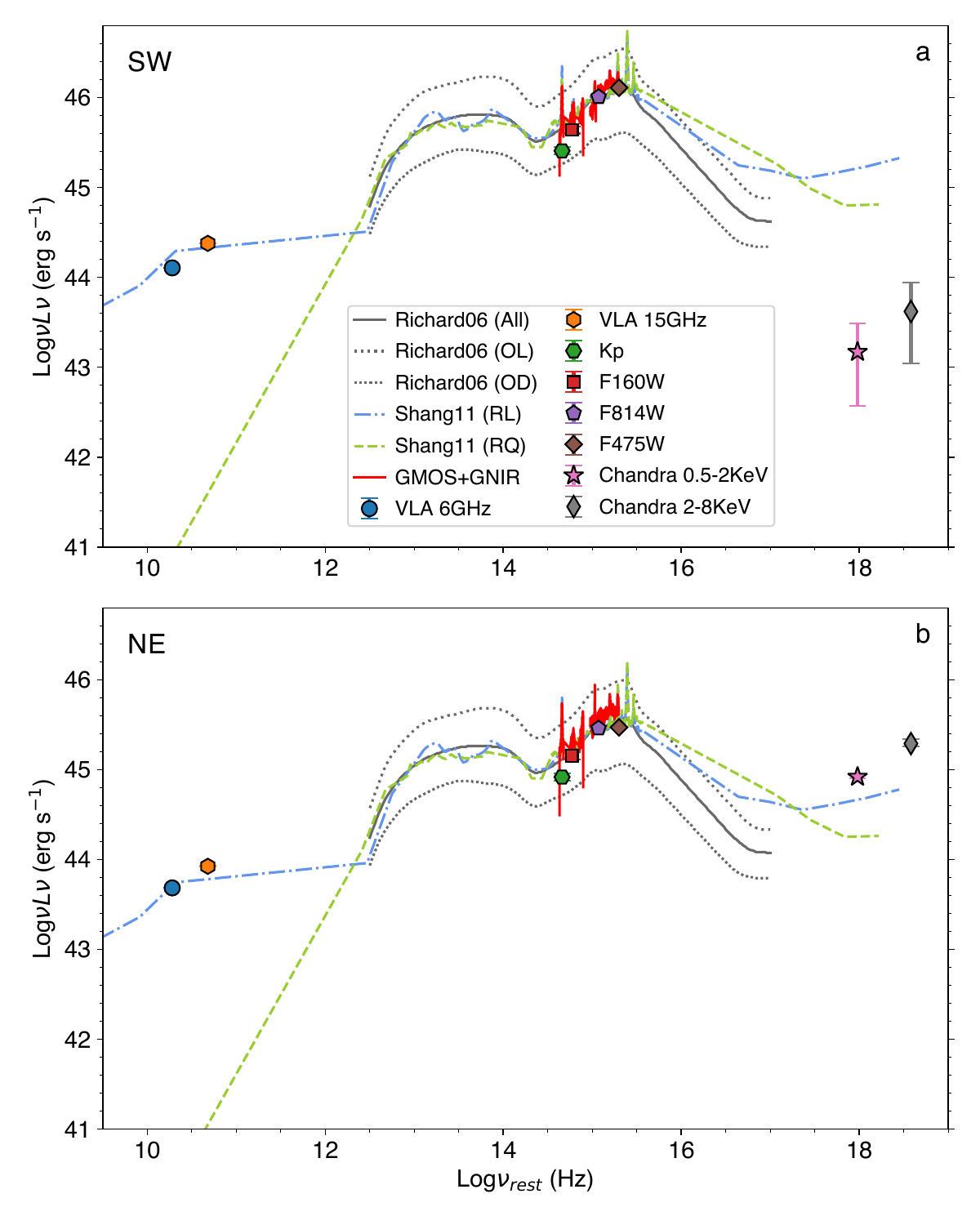}
    \caption*{
    \textbf{Supplementary Figure 10: Spectral energy distributions.} Panel a (Panel b): Multi-wavelength Spectral Energy Distributions (SEDs) of the SW (NE) nuclei in \obj . For comparison, the SEDs of typical optically selected SDSS quasars\cite{Richards2006} are shown in grey, with the mean values shown in solid (``All''), and the optically luminous (``OL'') and dim (``OD'') sub-populations shown in dotted curves. Also shown are the SEDs of optically bright, non-blazar quasars from \cite{Shang2011}, including both radio-loud (``RL''; shown in dash-dotted curves) and radio-quiet (``RQ''; shown in dashed curves) populations. All comparison SEDs are normalized to match the luminosities of \obj\ in \hst\ F814W. All error bars are 1$\sigma$. }
    \label{fig:sed}
\end{figure}

\begin{figure}
    \centering
    \includegraphics[width=0.8\textwidth]{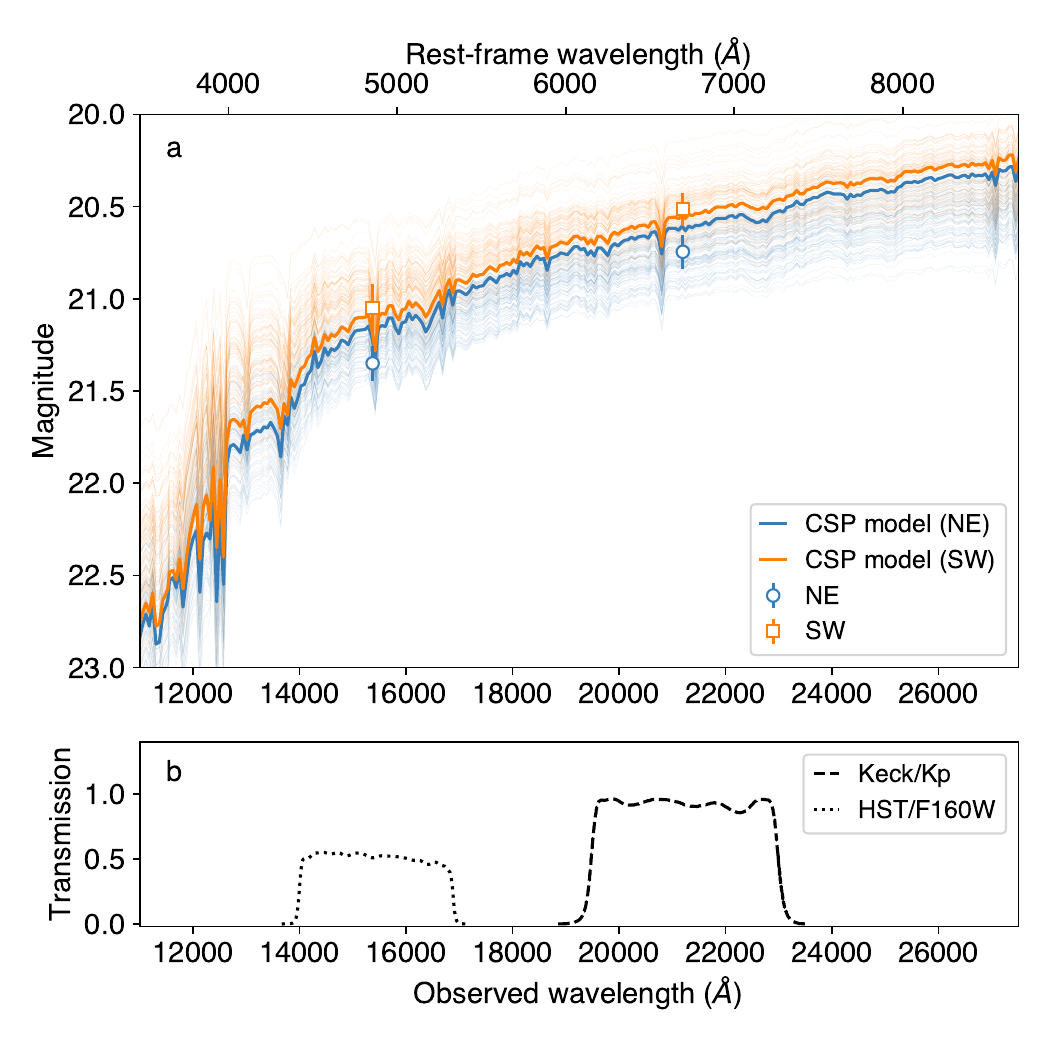}
    \caption*{
    \textbf{Supplementary Figure 11: Stellar population modeling. }
    Panel a: \hst\ F160W and Keck AO $K_p$ PSF-subtracted host-only photometry and the best-fit composite stellar population (CSP) models for the NE (in cyan) and SW (in orange) nuclei of \obj . The thick curves represent the best-fit CSP models whereas the thin ones are 100 random draws from 500k MCMC realizations (with 1k walkers and 1k steps with 500 burn-ins) for illustration. Panel b: Transmission curves for the \hst\ F160W (dotted) and Keck $K_p$ (dashed) filters, which roughly sample the rest-frame $g$ and $R$ bands for \obj. All error bars are $1\sigma$. }
    \label{fig:stellar_mass}
\end{figure}

\begin{figure}
  \centering
    \includegraphics[width=0.7\textwidth]{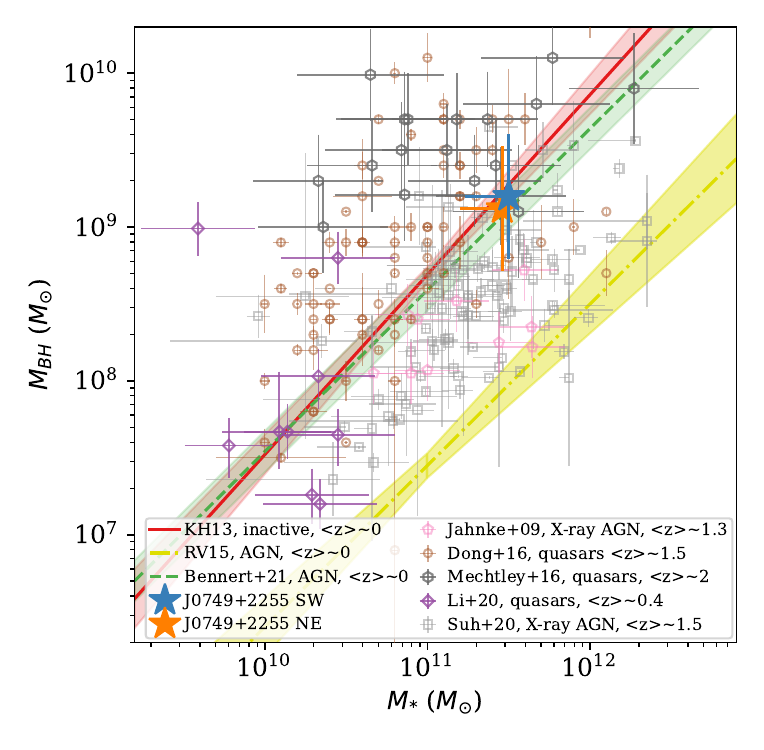}
    \caption*{
    \textbf{Supplementary Figure 12:
    Black hole mass versus host-galaxy total stellar mass.} The two nuclei in \obj\ are plotted as blue (SW) and orange (NE) stars with the error bars representing 1$\sigma$ uncertainties. Three local scaling relations are shown with a red solid line\cite{KormendyHo2013}, a green dashed line\cite{Bennert21}, and a yellow dash-dotted line\cite{Reines15}. The shaded regions denote the 1$\sigma$ scatters for these relations. Optical selected quasars at z$>$0.2 with HST IR images are shown as dark gray hexagons\cite{Mechtley16} and purple diamonds\cite{Li21}. X-ray selected AGN are shown as light grey squares\cite{Suh20} and light pink pentagons\cite{Jahnke09}. Quasars whose stellar masses are estimated from broad-band SED fitting (i.e., rather than spatially decomposing the PSF and the host) are shown as light brown circles\cite{Dong16}. All error bars are 1$\sigma$.
    }
    \label{fig:BH-Stellar_masses}
\end{figure}

\begin{figure}
  \centering
    \includegraphics[width=0.9\textwidth]{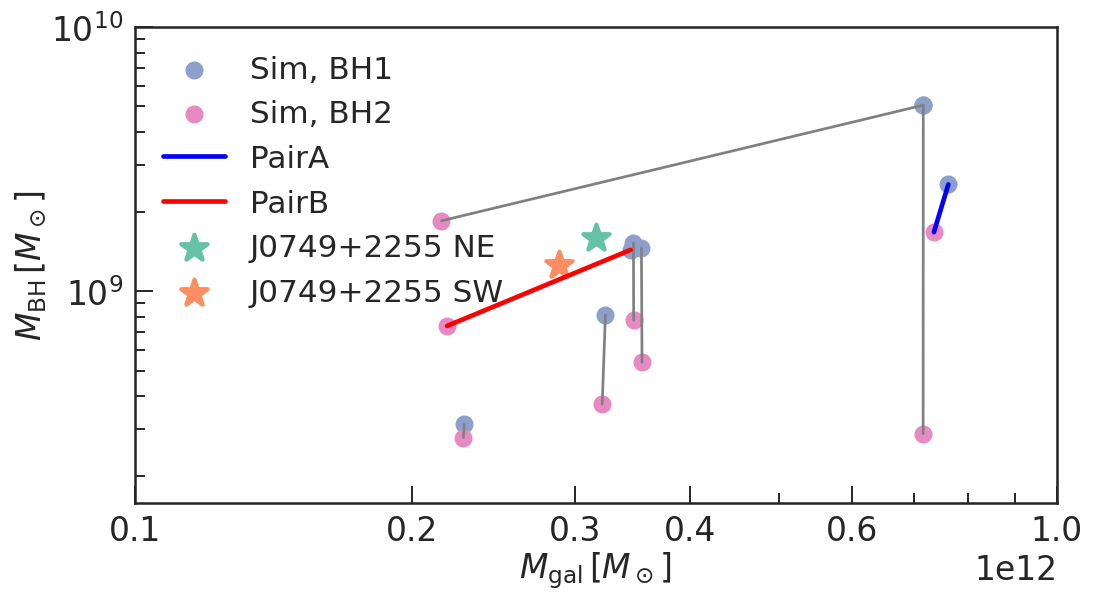}
    \caption*{
    \textbf{Supplementary Figure 13: Simulated black hole mass versue host stellar mass rerlation.} $M_{\rm BH}-M_{\rm gal}$ relation for the eight simulated bright quasar pairs in \texttt{Astrid}. The colored lines mark the two pairs (PairA and PairB) with apparently distinct galaxy bulges, although the kinematic bulge-disk classification adopted by simulations cannot be reliably used in interacting systems whose morphology is typically disturbed and the disks are in the process of being disrupted in the merger process. PairB is the closest match with \obj\ in both the SMBH masses and the host galaxy stellar masses.
    }
    \label{fig:SimMgal}
\end{figure}


\begin{figure}
  \centering
    \includegraphics[width=0.95\textwidth]{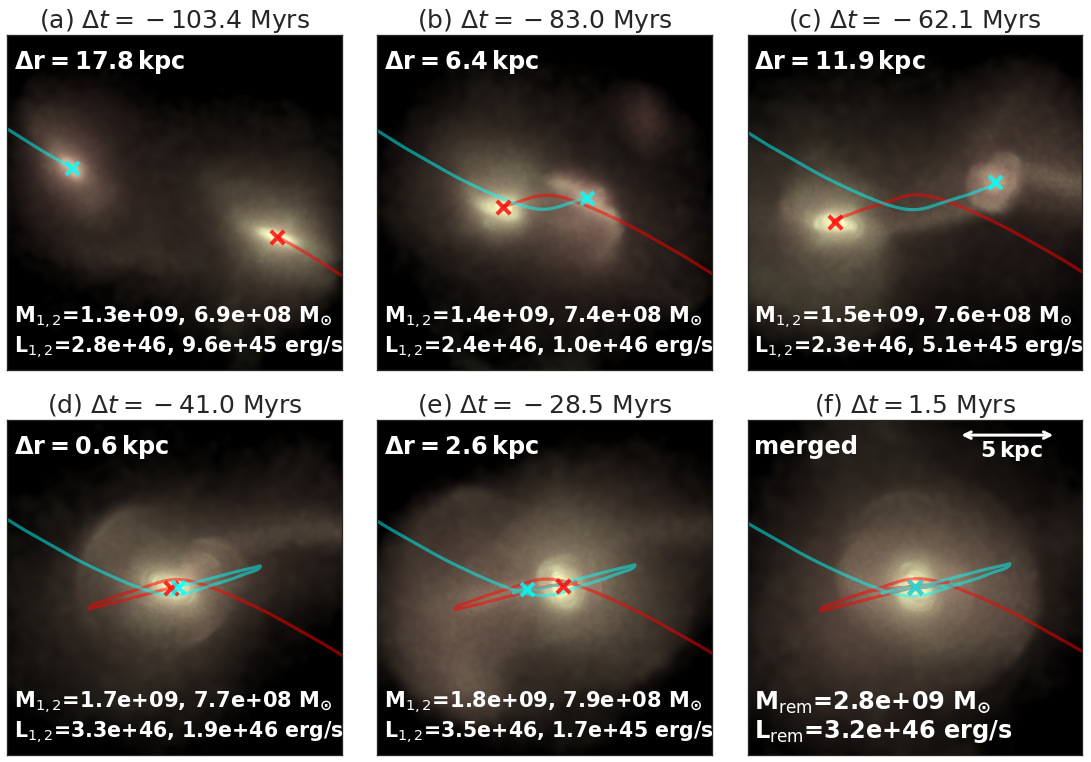}
    \caption*{
    \textbf{Supplementary Figure 14: Simulated dual quasars.} A simulated quasar pair (shown in red and cyan crosses) at $z\sim 2$ closely matching the properties of \obj , shown before and after the host galaxy merger.
    $\Delta t$ refers to the time difference between the snapshot and the SMBH merger. Panel b is similar to \obj\ at the time of observation, with slightly lower SMBH and galaxy masses. The two SMBHs merge in the simulation $83\,{\rm Myrs}$ after the ``observation'', and the merger remnant has $M_{\rm BH}\sim 3\times 10^9\,M_\odot$.
    }
    \label{fig:SimDual}
\end{figure}

\clearpage

\bibliography{refs}

\end{document}